\documentclass[onecollarge]{svjour2}
\usepackage{graphicx}
\usepackage{mathptmx}
\usepackage{amsmath}
\usepackage{amssymb}

\newcommand*{\citenum}[1]{\cite{#1}}
\newcommand*{\avg}[1]{\ensuremath{\left\langle{#1}\right\rangle}}
\newcommand*{\C}{\ensuremath{\mathcal C}}
\newcommand*{\F}{\ensuremath{\mathcal F}}
\renewcommand*{\H}{\ensuremath{\mathcal H}}

\newcommand*{\Sx}{\ensuremath{\mathbb S_x}}
\renewcommand*{\Pr}[1]{\ensuremath{\operatorname{P}\left({#1}\right)}}

\newcommand*{\pd}[2]{\ensuremath{\frac{\partial {#1}}{\partial {#2}}}}
\newcommand*{\eval}[2]{\ensuremath{\left.{#1}\right|_{#2}}}

\journalname{Journal of Statistical Physics}

\begin{document}

\title{ Information Theory and Statistical Mechanics Revisited}
\author{David M. Rogers \and Thomas L. Beck \and Susan B. Rempe}
\institute{David M. Rogers \and Susan B. Rempe \at
Center for Biological and Materials Sciences, MS 0895 \\
Sandia National Laboratories, Albuquerque, New Mexico 87185, USA\\
\email{dmroge@sandia.gov}
\email{slrempe@sandia.gov}
 \and
Thomas L. Beck \at
Department of Chemistry\\
University of Cincinnati, Cincinnati, Ohio 45221-0172 \\
\email{thomas.beck@uc.edu}
}


\maketitle

\begin{abstract}
  The statistical mechanics of Gibbs is a juxtaposition of subjective, probabilistic ideas on the one hand and objective, mechanical ideas on the other.  From the mechanics point of view, the term `statistical mechanics' implies that to solve physical problems, we must first acknowledge a degree of uncertainty as to the experimental conditions.  Turning this problem around, it also appears that the purely statistical arguments are incapable of yielding any physical insight unless some mechanical information is first assumed.  In this paper, we follow the path set out by Jaynes\cite{ejayn57}, including elements added subsequently to that original work, to explore the consequences of the purely statistical point of view.
Because of the amount of material on this subject, we have found that an ordered presentation, emphasizing the logical and mathematical foundations, removes ambiguities and difficulties associated with new applications.  In particular, we show how standard methods in the equilibrium theory could have been derived simply from a description of the available problem information.  In addition, our presentation leads to novel insights into questions associated with symmetry and non-equilibrium statistical mechanics.
Two surprising consequences to be explored in further work are that (in)distinguishability factors are automatically predicted from the problem formulation and that a quantity related to the thermodynamic entropy production is found by considering information loss in non-equilibrium processes.
Using the problem of ion channel thermodynamics as an example, we illustrate the idea of building up complexity by successively adding information to create progressively more complex descriptions of a physical system.  Our result is that such statistical mechanical descriptions can be used to create transparent, computable, experimentally-relevant models that may be informed by more detailed atomistic simulations.
We also derive a theory for the kinetic behavior of this system, identifying the nonequilibrium `process' free energy functional.  The Gibbs relation for this functional is a fluctuation-dissipation theorem applicable arbitrarily far from equilibrium, that captures the effect of non-local and time-dependent behavior from transient driving forces.  Based on this work, it is clear that statistical mechanics is a general tool for constructing the relationships between constraints on system information.
\end{abstract}


\section{ Introduction}

  If the foundation of thermodynamics is to be built on processes existing in the physical world, then the whole structure of the theory will be subject to constant revision as new physics is discovered.  This, however, has not proven to be the case.  Rather, as new mechanistic information is added, statistical mechanics persists in an identical form, with changes only in the meaning attached to system states and measurement outcomes.  It follows that, as in the case of the geometry of Euclid, statistical mechanics does not describe objects actually existing in the physical world, but rather idealizations of them\cite{hpoin07}.  This distinction immediately explains why the structure of statistical mechanics has persisted throughout the developments of the last century.  Because its basic axioms are conventions chosen to be logically consistent and in agreement with our intuition, the mathematical form of the theory operates as a device for carrying out extended logic.

  There have, to date, been many examples of using logical inference for framing
statistical mechanical questions.  Perhaps the most widely known is in the gradual
shift in the conceptualization of an ``ensemble."
Early ideas, associated with the names of Maxwell, Boltzmann, and others,
were based on physically realizable systems with many weakly interacting particles, {\em i.e.} gases.  The theory simply stated that an examination of all the particles at a single instant revealed the statistical properties of the ensemble. 
Gibbs~\cite{jgibb02} adapted the concept to systems that may contain strong
internal interactions, {\em e.g.} solids or condensed phases, by imagining the
ensemble as an infinite number of physical replicas of the system.  His subjective conceptualization can be seen in his definition of the laws of thermodynamics as expressing ``the approximate and probable behavior of systems of a great number of particles, or, more precisely, $\ldots$ for such systems as they appear to beings who have not the fineness of perception to enable them to appreciate quantities of the order of magnitude of those which relate to single particles, and who cannot repeat their experiments often enough to obtain any but the most probable results."  It was
immediately clear that for developing a subjective, formal treatment of the probability
distribution over phase and its consequences, `hypotheses concerning the constitution of
matter' would not be required except in working out special cases.

  The shift toward a subjective interpretation occurred only gradually because of
the combination of Gibbs' modest personality\cite{jmehr98} and a dispute between Gibbs and his
contemporaries~\cite{pehre59}, who viewed the physical reason for the weak
coupling between ensembles which brought about equilibrium as paramount.  Even
as Schr\"{o}dinger~\cite{eschr67} presented a maximum entropy derivation of the
canonical ensemble similar to the modern treatment of Jaynes\cite{ejayn03}, he
still found it necessary to seek a middle-ground by considering such distractions as the
physical realizability of infinite heat baths.  The work of
Jaynes~\cite{ejayn57,ejayn83,ejayn03} and others~\cite{wgran87} went a great
deal toward clarifying the situation by making a distinction between the
``delusion that an ensemble describes an `objectively real' physical
situation"~\cite{ejayn83} and the subjective question of determining the
``agreement between the premises and the conclusions."~\cite{jgibb02}  However,
the philosophical debate over objective vs. subjective interpretations of thermodynamics continues to date~\cite{nvank07}.
Not surprisingly, attempts to prove ergodicity and convergence to maximum entropy distributions using mechanical arguments show that the most robust route is to introduce some form of uncertainty\cite{rzwan60,mmack89}.

  Perhaps the strongest criticism of this approach is associated with the use of the term, `subjective.'  This term seems to imply that the results of the theory cannot be considered as objectively existing in reality.  Nevertheless, experiments are able to compare work and heat values to find agreement with thermostatics, provided a given system behaves according to the assumptions.  In exactly the same way, Euclid's geometry is able to deduce physically measurable distances, provided these objects behave as ideal solids.  Subjectivity is present in both of these cases because assumptions are always required in order to calculate one quantity from another.  The term `subjective' simply acknowledges that this reasoning process proceeds from assumptions derived from experience.  Physical predictions of objectively real phenomena can be made from a subjective theory based on assumptions that are objectively correct.  However, imputing objectivity to assumptions used to solve a particular problem makes it impossible to conceive of possible changes in prior information and has given rise to some of the most difficult paradoxes in science.
  
  In this paper, we aim to derive the statistical aspect of thermodynamics from the logical foundation given by Jaynes~\cite{ejayn03} in sufficient detail to present applications to modern problems outside the realm of the equilibrium canonical ensemble.  Although some of the most important results of this inquiry have already been presented by Gibbs, we find that a derivation from first principles clarifies the logical foundations of the theory.  A similar derivation for the canonical ensemble from first principles\cite{ejayn03} exemplifies the generality that such a theory may attain; however, several important questions remain unaddressed.  First and foremost, the form of the canonical ensemble must change when new degrees of freedom are added.  This addition corresponds to a change in the prior information for the problem, and it is not immediately evident how both problems can be related.  We have found that directly attacking this change in the number of possible `states' of a system requires a type of logical relativity theory, which results from rejecting the two-valuedness of elementary hypotheses.  We address this theory in Sec.~\ref{sec:logos} and provide the most important details in the Appendix.

  Next, we introduce the partition function and entropy functionals in Sec.~\ref{sec:info}, from which the elementary theory of statistical mechanics becomes manifest.  A relative entropy functional is in most respects simpler than the free energy as it may be deduced from purely statistical considerations.  The basis of this functional in information theory shows that thermodynamic states are characterized not by an objective physical situation, but instead by subjective information about the system.  Further, logically consistent predictions from statistical mechanics will disagree with the results of experiment whenever physically incorrect assumptions are made about the state of the system.  A novel result of this section is that thermodynamic free energies fundamentally express the log-likelihood of a state of knowledge.  Although the absolute probability is undefined without specifying all possible states, the likelihood ratios between states may be deduced, and are exactly the exponentials of free energy differences.
Although the partition function and entropy are defined as `state-functions,' dependent on a state of knowledge, we justify the fundamental importance of likelihood ratios by confirming that the partition function can be built from a product of such likelihood ratios in any order, leading to the conception of thermodynamic cycles.  This result also proves that the resulting probability distribution is a state function, as it should be.
The assumptions required for constructing the relationships between informational states--the laws of statistical mechanics--are thus founded on probability theory.

  We will develop the ion channel as an example problem for applications (Sec.~\ref{sec:apps}) of the basic set of relations given in Sec.~\ref{sec:info}.  Transmembrane proteins that to shuttle solutes between two aqueous/membrane interfaces have drawn the attention of a large crowd of experimental and theoretical investigators.\cite{bhill01}  Selective channels and transporters are critical for maintaining living cells in their nonequilibrium state.  Similar functionality is a required ingredient of synthetic semi-permeable partitions, used in fuel cells, solute separation, and electrochemical sensing.  The operational characteristics of these devices are determined from their response to applied pressure, electric fields, and solute concentration differences.  The most easily measured response is ion conduction, available through current measurements that can be carried out on micrometer-sized patches at milli-second resolution.\cite{ohami81,wwond90}  Conduction of other species, such as water, as well as structural changes in the channel and surrounding interface regions are also important, but less accessible.  The most easily accessible theoretical descriptions of channel behavior center around the structural properties of the equilibrium state and its propensity for ion occupancy under no external bias (in non-conducting conditions).\cite{broux04}  Instead of presenting a patchwork of accumulated techniques in statistical mechanics, in this article we present a top-down view by successively adding mechanistic information to predict these propensities.  This allows a construction of the simplest possible physical interpretation of channel behavoir, but uses a statistical mechanics capable of deriving all the complexities of atomistic and quantum-mechanical systems.  Because no net currents are present at equilibrium\cite{ejayn86}, the fluxes in these systems must be analyzed using a nonequilibrium theory.

  The usual Komologrov definition of probability and the Boltzmann factor are developed in Sec.~\ref{sec:constr} and~\ref{sec:maxent}.  The latter is derived by a maximum relative entropy argument along a path in a thermodynamic cycle.  For the special case of maximum relative entropies, we find that the entropy increments add to the total information entropy of each state, proving path independence of the entropy for this case.  These two cases generate the usual equilibrium thermodynamics approach without the necessity of assuming extensivity.  We will show that this approach can be used directly to give a na\"{\i}ve distribution over ion occupancy states for the channel.  We will show later that this distribution is related to a coarse-grained (marginal) distribution at a state of knowledge with more degrees of freedom.  However, both states of knowledge correctly employ the rules of statistical mechanics, and their difference lies in the assumed information for the problem.

  Adding new coordinates, the opposite of restricting them in Sec.~\ref{sec:constr}, leads to the multicanonical ensemble, presented in detail in Sec.~\ref{sec:multi}.  The new degrees of freedom are termed coarse coordinates.  The relationship between canonical and multi-canonical ensembles is the usual one.  Fixing coarse coordinates within the multicanonical ensemble generates a conditional ensemble.  The aggregate probability of the coarse coordinates is related to the potential of mean force as in coarse-graining.\cite{jkirk35}  Although we could directly add time-dependent states in this section, we have adopted a slower development, adding conformational states of the channel at equilibrium.  This allows an intuitive connection to the well-known equilibrium theory, where introduction of interacting systems can change the distribution over the system of interest.  The analogous development in Ref.~\citenum{ejayn57} is the derivation of the constant pressure or constant angular momentum ensemble.  Instead of describing everything in terms of all atomistic positions and momenta, or all atomic and electronic eigenfunctions, we have shown here that these limits can be approached incrementally as necessary for each application.

  In general, adding maximum relative entropy information along with new coordinates, $Y$, will change the distribution over the previous coordinates, $X$.  In some cases, for example when the marginal distribution over $X$ is experimentally known, this is not desirable.  We instead seek a method for inference on $Y$ from a known distribution over $X$ and maximum entropy information.  The method for constrained addition is derived in Sec.~\ref{sec:cond} by assuming a probability distribution for ion occupancy states in a channel, and then inferring the distribution of channel conformations.  Other important applications of the same theory are possible.  In particular, it leads directly to the predictive statistical mechanics\cite{ejayn86} of dynamic processes arbitrarily far from thermodynamic equilibrium.  The non-equilibrium process entropy (caliber) and free energy functionals follow naturally from a specification of non-equilibrium states as trajectories.  The Gibbs relations for these functionals lead trivially to generalized fluctuation-dissipation theorems.  Although based on ideas originally from Jaynes\cite{ejayn79,ejayn80}, the work presented in this section differs in an important respect.  By fixing the distribution at an initial time and then maximizing the step-wise {\em transition entropies} we arrive at a non-anticipating process.  Previous presentations\cite{ejayn79,cmaes99,rdewa03} utilize anticipating conditions, resulting in the possibility of non-physical influences from forces which may be exerted on the system at future times.  Removing this shortcoming brings us closer to an original idea by Jaynes\cite{ejayn57a} and gives path probabilities immediately recognizable as a canonical form for forward transition processes.  Information loss in discarding the starting distribution in favor of the final distribution leads to a quantity analogous to the thermodynamic entropy production.  This entropy has a great advantage over other formulations\cite{gcroo00,etrep04} in that it does not explicitly require definition of a `steady-state.'  Such a state may not be unique (as in the case for the Liouville equation) or even exist, e.g. transient processes like evaporation in an open system.  It is our hope that this new development will complete the statistical foundations of thermodynamics by providing a basis for the second law of thermodynamics in information theory already hinted at in the problem of Maxwell's demon.\cite{hleff03}

\section{ Logical Foundations}
\label{sec:logos}

  Jaynes\cite{ejayn03} presents a cogent interpretation of probability theory as a method for conducting logical inference in the presence of uncertainty.  This interpretation is based on P\'{o}lya's qualitative conditions for plausible reasoning in mathematics\cite{gpoly54} combined with the consistency theorems of Cox and Acz\'{e}l\cite{rcox61,jacze87} deduced by consideration of the associativity equation.  Requiring our system for assigning plausibilities to be associative, such that adding information in any order leads to the same probability assignment, it is possible to deduce the product rule
\begin{equation}
\label{eq:Bayes}
\Pr{AB|C}=\Pr{A|BC}\Pr{B|C}=\Pr{B|AC}\Pr{A|C}
,
\end{equation}
for which the right equality is Bayes' theorem.  The symbols, $A,B,\text{ and }C$ stand for hypotheses, or logical propositions, and the symbols on the right of the $|$ represent given information, or assumptions.  In this paper, we denote propositions using Greek or capital letters.  This distinction is necessary to allow for propositions that represent coordinates, i.e.
\begin{quote}
$X$: Some property of the system is described by the number $x$.
\end{quote}

  Propositions always appear inside the probability symbol and follow the Boolean algebra, where multiplication denotes a logical `and,' while addition represents a logical `or.'  We refer the reader to the first few paragraphs of the Appendix for the necessary notation.  Of particular importance is the relation $AB=A$ when $A\Rightarrow B$, used extensively to replace $X$ with $X\Sx$.  We also omit the prior information $I$ for clarity in some instances, although it is always to be assumed in the formulas presented here.

  An immediate question occurs as to how probabilities may be assigned in the first place.  The most appealing answer is to employ the principle of indifference (termed $I$), which states that, for any number of possible outcomes, each is equally probable.  However, this again begs the question as to the definition of the hypothesis space.  Is any assignment possible in the absence of this knowledge?  We assume that some assignment is possible, and state it as $\Pr{A|I}=\text{constant}$ for hypotheses, $A$, that are `of the same type.'  We provide a formal justification for this process in the Appendix, and note that it extends some amount of inference to statements that are undecidable, but does not affect the conditional assignments, $\Pr{C|AI}$, when $A$ says something about $C$.

  The ability to reason in an un-defined hypothesis space has some interesting consequences for the principal of complementarity.\cite{ejayn89}  Suppose an infant is entertained by a screen that shows one color, $x_1,x_2,\text{ or } x_3$ and plays one sound, $y_1,y_2,\text{ or }y_3$ at every moment.  However, $X_1$ never occurs simultaneously with $Y_1$ and so on for $X_2\text{ and }Y_2$, $X_3\text{ and }Y_3$.  As these ideas are being learned, it appears that $X_1$ and $Y_1$ are mutually contradictory circumstances, and it should be possible for the infant to express some idea of the validity of that statement, $X_1\oplus Y_1$.  Upon their first encounter with both $X_1Y_1$ in the real world, the infant may be genuinely surprised.  Moreover, if $X_1$ represents the statement, `color $x_1$ is present,' and the child believes all colors are mutually exclusive ($\oplus(X_1,X_2,X_3)$), then $\bar X_1$ can be represented equally well with, `any color other than $x_1$ is present.'  It may thus be ontologically true that both $X_1\bar X_1$ for a complex scene.\cite{chamb67,kbimb01}  Following the argument of Poincar\'e for establishing real numbers\cite{hpoin07}, suppose the screen is divided into smaller and smaller segments, and each time the colors are found to be mutually exclusive in each segment.  Then the act of defining, by recursion, a continuous space on which colors are mutually exclusive will lead to an idea in contradiction with the nature of light (but not with our measuring apparatus).  The situation is immediately seen to be similar to the double-slit experiment, where we must abandon our notion that a particle at position one, $X_1$, and at position two, $X_2$, are mutually exclusive, and instead replace it with $X_1+X_2+(X_1\oplus X_2)C$, where $C$ denotes `a measurement has been taken to determine $x$.'  These considerations relax the `logical consistency' restrictions on what questions may be asked of the quantum theory.\cite{romne88}

  The problem of assigning relative probabilities to hypotheses concerning which sets of events may be possible is considered at length in the Appendix.  Using the principle of indifference, the main result is that the probability of a set of independent events, $\Omega$, is proportional to the number of events, $|\Omega|$, so that 
\begin{align}
\Pr{x|\Omega I} &= \frac{1}{|\Omega|}, \; x\in\Omega \notag \\
                &= \frac{\Pr{x|I}}{\Pr{\Omega|I}} = \frac{\text{const.}}{\Pr{\Omega|I}} \notag \\
\label{eq:pom}
\Rightarrow & \Pr{\Omega|I} \equiv |\Omega| \Pr{\varphi|I}
,
\end{align}
where $\varphi$ stands for some elementary hypothesis.
This result elegantly sweeps questions due to symmetry under the rug, and these will be given more consideration in a subsequent paper.  The discussion in the Appendix already shows that this development provides a method for dealing with symmetric hypotheses in a very simple format, fundamentally based on the principle of indifference.  The appropriate `(in)distinguishability factors' are derived as a result of its use in Sec.~\ref{sec:constr}.  From a statistical mechanics viewpoint, the principle of indifference then provides partition functions, $Z[\Omega] = \Pr{\Omega|I}/\Pr{\varphi|I}$, consistent with completely `entropic' systems.  We show next that conventional partition functions may be obtained from these by moving constraints on average values to the left-side of the probability symbol as well.

\section{ Minimal Relations of Statistical Mechanics}
\label{sec:info}

  Because we are deriving purely statistical relationships, the only things we are able to compare are states of knowledge.  There may be several convenient computational notations or methods for solving the resulting equations, but it will not be necessary to read these as implying physically existing quantities or mechanisms.  It is not the physical, causal mechanisms themselves, but rather theories about them that are the subject of the reasoning process.  These theories may appear as propositions to be mutually compared or as given information for solving certain inference problems.  However, unless physics appears in this way, it can have no influence on the solution of the logical problem.  Mechanics enters because the set of coordinates and constraints relevant to any given hypothesis must be found from mechanical insight, and the answers resulting from statistics will depend non-trivially on this input.

  We claim that the state functions of statistical mechanics, the partition function and entropy functional, can be derived by successive addition or replacement of problem information.  The first process can always be carried out, with a corresponding change in the probability distribution for system states {\em via} re-weighting the probability distribution from the previous state.  In general, this process is uni-directional.  The second process, replacing information, can only be carried out directly {\em via} re-weighting in certain circumstances.

  To develop our notation, we represent each state of knowledge by a set, $\C = \{A\}$, of propositions or informational constraints, $A$.  Important types of propositions include system coordinates, energy assignments, and statistical weights.  As we will see, the latter amount to propositions of the type ``There is a physical mechanism increasing the likelihood of state $x_1$ over $x_2$ by some amount,'' and are closely aligned with energy assignments and the translation of problems with pre-specified coordinates to problems with pre-specified physical forces.  In addition to these, we also permit statements defining the set of coordinates relevant to deciding a given proposition--the problem phase space--as well as the symmetries these propositions obey.  Although it may seem peculiar from a mechanistic point of view, any problem constraints can appear either as given information or as objects to be compared.  We might also be able to remove the distinction between coordinates, energy assignments, and definitions of phase space to form so-called generalized ensemble or coarse-grained systems.

  We consider the process of adding information $A$ to some known state, $\C$, which we denote by $\C\to A\C$.  The first quantity of interest is the probability $\Pr{A|\C I}$, which we compare to an alternative process, $\C\to\Phi\C$.  It is convenient to assume the existence of a null hypothesis, $\Phi$, that is un-decidable from any other information.  Formally,
\begin{equation}
\label{eq:null}
 \Pr{\Phi|\C I} = \Pr{\Phi|\C'I} = \Pr{\Phi|I}\; \forall \C,\C'
.
\end{equation}
Now divide the set of propositions, $\C'$ (appearing above), into two sets, $\mathcal D$ and $\C$.  From the two equivalent ways of composing $\Pr{\mathcal D \Phi|\C I}$ using Bayes' theorem, it is easy to see that the above is true if and only if $\Phi$ is irrelevant to conclusions about $\mathcal D$. 
\begin{equation*}
 \Pr{\mathcal D | \Phi \C I} = \Pr{\mathcal D | \C I}
\end{equation*}

  We compute the relative likelihood,
\begin{equation}
\label{eq:Z1}
Z[A\C]/Z[\C] \equiv \frac{\Pr{A|\C I}}{\Pr{\Phi|\C I}} = \frac{\Pr{A\C|I}/\Pr{\Phi|I}}{\Pr{\C|I}}
,
\end{equation}
using
\begin{equation}
\label{eq:LR}
\frac{\Pr{A|\C I}}{\Pr{\Phi|\C I}} = \sum_{\{X\}} \frac{\Pr{A|X\C I}}{\Pr{\Phi|X\C I}} \Pr{X|\C I}
.
\end{equation}

  The summation set, $\{X\}$, should include any system states relevant to deciding the plausibility of $\C$ or $A$.  To see this, assume that the states relevant to deciding $\C$ or $A$ are collected in the space $\Sx$.  Then write $\{X\}=\Sx\times\mathbb S_i$, where $X_i\in\mathbb S_i$ are irrelevant to $A$ and $\C$ so that $\Pr{X|A\C I}=\Pr{X_{A\C}X_i|A\C I}=\Pr{X_{A\C}|A\C I}\Pr{X_i|I}$.  The sum in Eq.~\ref{eq:LR} factors into
\begin{equation*}
\sum_{X_{A\C}\in\Sx}\sum_{X_i\in\mathbb S_i} \frac{\Pr{X_{A\C}X_iA|\C I}}{\Pr{\Phi|\C I}}
  = \sum_{X_{A\C}\in\Sx} \frac{\Pr{A|X_{A\C}\C I}}{\Pr{\Phi|X_{A\C}\C I}} \Pr{X_{A\C}|\C I}
.
\end{equation*}

  If we are use information $A$, as an assumption it should come from known experimental data on the system.  In order to establish $A$, we may therefore tabulate frequencies for $X\in\Sx$.  If $A\C$, turned out to be true, scientists basing their conclusions only on $\C$ would be increasingly surprised (or skeptical if the report is second-hand) at the evidence collected after $N$ trials.  This is because the probability of these results given $\C\Sx$ would be (from the multinomial distribution),
\begin{align}
 \Pr{\{X\}_1^N\sim A\C|\C\Sx} &= N! \prod_{X_i\in\Sx} \Pr{X_i|\C\Sx}^{n_i} / n_i! \notag \\
  &\to e^{N\H[A\C\Sx|\C\Sx]} \notag \\
\label{eq:H}
\H[A|B] &\equiv -\sum_{X_i\in\Sx} \Pr{X_i|A\Sx}
               \ln \left[ \frac{\Pr{X_i|A\Sx}}{\Pr{X_i|B\Sx}} \right]
.
\end{align}
According to $\C$, the likelihood of such a set of observations decreases exponentially with N.  This is a condensed version of the Wallace derivation for the entropy, presented in more detail in Ref.~\citenum{ejayn03}.  The limit taken in the second equation is as $N\to\infty$, which is appropriate for assessing such a set of hypothetical observations or second-hand reports.  Evidently, the Kullback-Liebler divergence, $-\H \ge 0$, represents the value of the information $A\C$ (or difference of opinion) to an observer who has already accepted $\C\Sx$.  The relative information entropy, $\H$, reaches its maximum, zero, when the new information does not alter the distribution.  For any reasonable comparison to be made, the distributions must be compared over the same set, $\Sx$, which should include any observational information that $A$ or $B$ may predict.  As in the case for the free energy difference, above, the relative entropy is independent of the distribution over irrelevant variables, $X_i\in\mathbb S_i$.  This happens here because the probability assignments are identical over the subspace $X|X_i$ for each $X_i$.

\subsection{ Transitivity}

  The above arguments showed how to add information incrementally.  If the starting point is taken to be $I$, it is then possible to assign a rank to all states,
\begin{align}
\label{eq:Z}
Z[\C] &\equiv \sum_{\{X\}} \frac{\Pr{\C|X I}}{\Pr{\Phi|X I}^{|\C|}} \Pr{X|I}
,
\end{align}
where $|\C|$ denotes the size of the set, $\C$, and $Z[\Phi]=1$.
And we may compare any two states using
\begin{equation}
\label{eq:DF}
\frac{\Pr{\C| I}}{\Pr{\mathcal D| I}} = \frac{Z[\C]}{Z[\mathcal D]}
.
\end{equation}

  To show that this can be computed using successive addition of information and Eq.~\ref{eq:Z1}, it is necessary to prove that any order of information addition leads to (\ref{eq:Z}).  The proof is a direct consequence of Bayes' theorem (\ref{eq:Bayes}).
\begin{align*}
 \Pr{AB|I} &= \Pr{A|I} \Pr{B|AI} \\
  &= \Pr{A|I} \sum_{X\in\Sx} \Pr{BX| AI} \\
  &= \Pr{A|I} \sum_{X\in\Sx} \Pr{B |XAI} \Pr{X|AI}
\end{align*}
The last formula is exactly the form of Eqns.~\ref{eq:Z1} and~\ref{eq:LR} (with the normalization removed) and the above derivation is symmetric in $A$ and $B$.

  The relative entropy cannot be so defined, since it compares distributions.  However, the entropy with respect to a complete space,
\begin{equation}
\label{eq:Hstate}
\H[\C\Sx|\Sx],
\end{equation}
can be compared among all states which depend only on the space $\Sx$.  In the sections below, it will be shown that adding maximum entropy-type information, $B$, makes $\H[AB\Sx|\Sx]=\H[AB\Sx|A\Sx]+\H[A\Sx|\Sx]$.  However, this is not necessarily true when Eq.~\ref{eq:Fadd} does not hold.

\subsection{ Inference}

  The above concepts may be solidified using the inference process as an example.  Given a model, $M$, for how data may be generated, we may use any prior information or symmetries of the problem to write down a prior state of knowledge, $\mathbb S_\theta M$.  The prior distribution over the parameter space, $\Pr{\theta|\mathbb S_\theta M}$, is then given by the free energy for the process $\mathbb S_\theta M\to\theta\mathbb S_\theta M=\theta M$.  Next, some data, $D_1$, is collected and the state of knowledge updated to $D_1\mathbb S_\theta M$.  
The free energy for $D_1\mathbb S_\theta M \to \theta D_1 M$ now gives the posterior distribution.  Bayes' theorem appears as the thermodynamic cycle identity between the free energy for
$D_1\mathbb S_\theta M \to \theta D_1 M$ and $D_1\mathbb S_\theta M \to \mathbb S_\theta M \to \theta M \to \theta D_1 M$
\begin{equation*}
\frac{Z[\theta D_1 M]}{Z[D_1\mathbb S_\theta M]} = \left(\frac{Z[D_1\mathbb S_\theta M]}{Z[\mathbb S_\theta M]}\right)^{-1} \frac{Z[\theta M]}{Z[\mathbb S_\theta M]}\frac{Z[\theta D_1 M]}{Z[\theta M]}
\end{equation*}
Interestingly, inference using Bayes' theorem is increasingly being used to estimate the probabilities of free energies from computational sampling experiments\cite{ssrir05,apoho06,droge08}, and this can be generalized to estimating free energy functionals.\cite{ghumm05,droge10a}

  The relative entropy between $\mathbb S_\theta M$ and $D_1 \mathbb S_\theta M$ measures how informative $D_1$ is in determining $\theta$, while $\H[D_1 D_2 \mathbb S_\theta | D_1 \mathbb S_\theta]$ shows the amount of information that $D_2$ conveys once $D_1$ is known.

\subsection{ Replacement of Information}

  The second type of process is that of completely replacing information.  For this case, we may apply all the formulas of the previous section, and compare $\C\to A\C$ with $\C\to B\C$.  However, instead of computing each of these separately, we wish to directly compare the total likelihood between $A$ and $B$.  We thus fix either situation, and use
\begin{equation}
\label{eq:direct}
1 = \sum_{X} \Pr{X|A \C} = \frac{\Pr{B| \C}}{\Pr{A| \C}} \sum_{X}
                                      \frac{\Pr{A|X \C}}{\Pr{B|X \C}} \Pr{X|B \C}
.
\end{equation}

  This shows that the distribution over $X|A$ can be transformed from $X|B$ {\em via} re-weighting -- although this is known to be computationally inefficient.\cite{gtorr77}  However, if there is an $X$ for which $\Pr{X|B \C}$ is zero, but for which $\Pr{A|X \C}=\tfrac{\Pr{X|A \C}\Pr{A|\C}}{\Pr{X|\C}}$ is non-zero, then the above expression cannot be evaluated.  Therefore, if $B$ contains a restriction on the set of allowable $X$, then this restricts mutual comparison among $A,B$.  Likelihood ratios can only be computed directly using Eq.~\ref{eq:direct} if $\Pr{X|B \C}$ is nonzero on a smaller space $\Omega\subseteq\{X\}$ than $\{X\}$ on which $\Pr{X|A \C}$ is nonzero.  To some extent, this caveat explains the computational problems involved with re-weighting samples.\cite{nlu01}

\begin{figure}[htbp]
   \centering
   \includegraphics[width=0.9\textwidth]{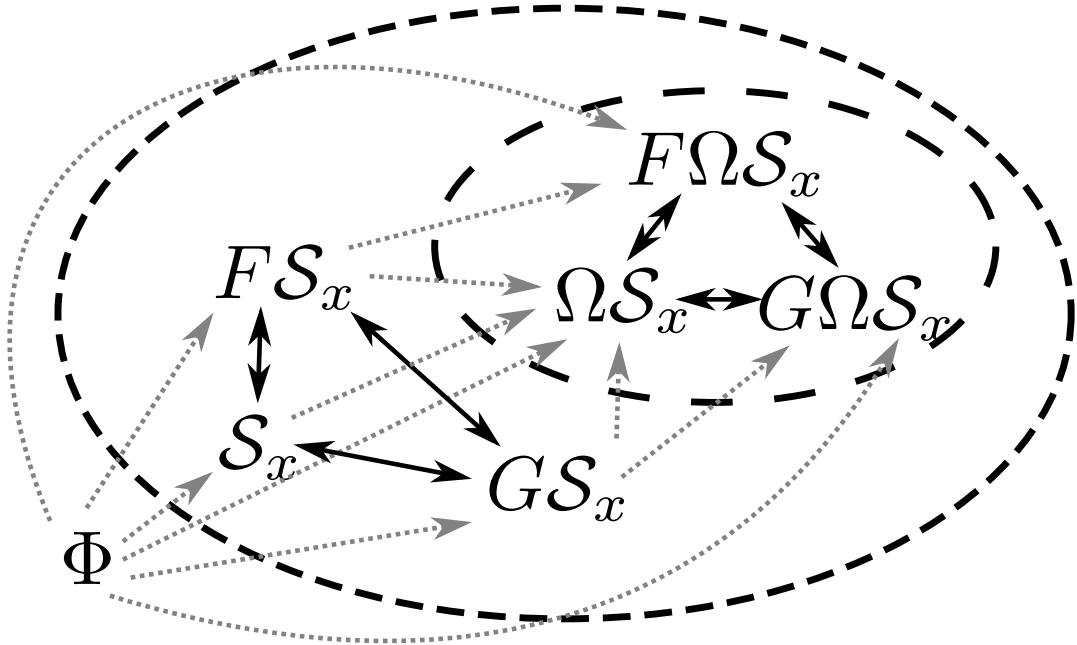} 
   \caption{Reaction diagram showing system states as nodes.  Two constraints, $\Sx$, defining a coordinate space, and $\Omega$, defining some further restriction are illustrated here.  $F$ and $G$ are average value constraints, and their relative likelihoods can be calculated using Eq.~\ref{eq:rewt} in either direction.  For identical constraints, all hypotheses are completely connected, as shown by the double-headed, dark arrows.  Restrictions such as $\Sx$ or $\Omega$ limit the set of propositions that can be directly compared without knowledge of $\Pr{\Omega|M}/\Pr{\Phi|M}$, and only one comparison direction is allowed, illustrated by the grey, dotted arrows.}
   \label{fig:rxn}
\end{figure}

  Propositions defined inside $\Omega$ can still be compared against one another, and their likelihoods computed from either the null hypothesis, $\Phi$, or a new null hypothesis, $\Phi\Omega$, defined relating only to $X$ allowed by $\Omega$.
Addition of the information, $B=B\Omega$, to a thermodynamic state can be represented using a commutation diagram (Fig.~\ref{fig:rxn}), where paths represent step-wise addition of constraints / hypotheses.  Completely commuting classes share an underlying definition of coordinate space.  Whenever information of the type $\Omega$ is added, it directly bears on subsequent propositions.  Paths adding $B\Omega$ will therefore restrict the set of subsequent questions that may be asked without knowledge of $\Pr{\Omega|\C}$.  These paths are therefore represented by a directed edge, branching from the above completely connected graph.  The commutation diagram terminology is justified by noting that the multiplicative functions, (\ref{eq:wt}), transforming one probability distribution into another arrive at the same distribution function for any `allowed' path.

  Because the free energy formula (\ref{eq:LR}) is simply Eq.~\ref{eq:direct} for the special case $B=\Phi$, it is convenient to define
\begin{align}
\label{eq:wt}
w_{B\to A}(X\C) &\equiv \frac{\Pr{A|X\C}}{\Pr{B|X\C}}, \\
\intertext{so that free energy differences can be expressed more simply as}
\label{eq:rewt}
\frac{\Pr{A|\C}}{\Pr{B|\C}} &= \avg{w_{B\to A}|B} .\\
\intertext{As their name implies, these are weights,}
\Pr{X|AI} &= \frac{w_{B\to A}(X) \Pr{X|BI}}{\sum_{X\in\Sx} w_{B\to A}(X) \Pr{X|BI}} \notag \\
\avg{f(x) | A} &= \frac{\avg{w_{B\to A} f(x)|B}}{\avg{w_{B\to A}|B}}
.
\end{align}
It must be understood that the re-weighting is only valid when $w_{B\to A} < \infty$.

\section{ Specific Applications}
\label{sec:apps}

  The formulas derived in the last section are well-known relations in statistical mechanics.  When Boltzmann factors are inserted for the weights $w_{A\to B}(x)$, Eq.~\ref{eq:rewt} generates free energy perturbation and umbrella sampling formulas\cite{fe07} and unambiguously identifies $\Pr{X|A}$.  However, a few important differences from the standard development can be noticed in the above.  First, the commutativity of thermodynamic cycles is perhaps not as widely appreciated as it should be.  Although it is well known that $Z[\C]$ is a state function, because of its definition in Eq.~\ref{eq:Z}, this shows that a sum of {\em relative} free energy differences around any closed loop of a thermodynamic cycle totals to zero with the caveat that Eq.~\ref{eq:rewt} may only be applied from a larger phase space to a smaller.  The same is not true of relative entropies (\ref{eq:H}), which give a sum dependent on the path taken.  Instead, it is necessary to define $\H[\C\Sx|\Sx]$ as the state function.  Also, the entropy definition of Eq.~\ref{eq:H} is independent of changes in phase-space volume because $\Pr{X|AI}$ transforms the same way as $\Pr{X|BI}$ for an injective change of variables $X\to Y$.
  
  The physical problem of determining $\Pr{A|X\C}/\Pr{B|X\C}$ has not yet been addressed.  Because this function can be expressed as a ratio, we need only specify $w_{\Phi\to A}(X\C)$.  Extending the concept of the partition function (Eq.~\ref{eq:Z}), the weights can be interpreted as $w_{\Phi\to A}(X\C) = Z[AX\C]/Z[X\C]$.  We will present arguments for defining this function for several different types of problems, and find that the standard Boltzmann-factor form, $e^{-\beta_A U_A(x)}$, is not a universal answer.  The general idea will be to find a minimal set of relevant information $XY$, implied by $X\C$ so that $A$ is conditionally independent from $\C$ when $XY$ is known, simplifying the weight to $w_{\Phi\to A}(X\C)=w_A(XY)$.  Comparing $w_A$ for different $XY$ then suggests an appropriate relative weight.  Specific problems relating to changes in the symmetries of phase space will be addressed in a separate paper.

\subsection{Constraints on Phase Space}
\label{sec:constr}

  A simple type of constraint is one that limits hypothesis space.
\begin{quote}
$\Omega$: The set of allowed states is limited to those in which $\C$ is a member of the set, $\Omega$.
\end{quote}
This type of constraint can be used to limit investigations to interesting, or highly probable configurations as well as formulate decision problems.
Adding $\Omega$ to a state results in a normalization
\begin{equation}
\label{eq:Onorm}
\Pr{\C|\Omega A} = \frac{\Pr{\C|A} I(\C\in \Omega)}{\sum_{\{\C\}} \Pr{\C|A} I(\C\in \Omega)}
,
\end{equation}
where the indicator function, $I(\cdot)$ is one when the condition is satisfied, and zero otherwise.

  Given some $\C$, two constraints that both allow $\C$ should be equally likely, leading to the assignment
\begin{equation}
\label{eq:wom}
w_\Omega(A\C) = \frac{\Pr{\Omega|\C A}}{\Pr{\Phi|\C A}} = I(\C\in \Omega)
,
\end{equation}
for any $A$ that does not specifically reference $\Omega$.

  The free energy of the constrained system is the denominator of Eq.~\ref{eq:Onorm}
\begin{equation}
\label{eq:dZconstr}
Z[A\Omega]/Z[A] = \sum_{\{\C\}} \Pr{\C|A} I(\C\in \Omega)
.
\end{equation}
This is consistent with the assignment in the appendix derived for the case $A=\varphi$ (Eq.~\ref{eq:pom}).

  Given two constraints, we can use Bayes' theorem to show
\begin{equation}
\label{eq:constr}
\Pr{\Omega_1|\Omega_2 F I} = \frac{\Pr{F\Omega_1\Omega_2|I}}{\Pr{F\Omega_2|I}}
  = \frac{\Pr{F|\Omega_1\Omega_2I}}{\Pr{F|\Omega_2I}} \Pr{\Omega_1|\Omega_2 I}
,
\end{equation}
a simple theorem relating free energies in successively constrained spaces.  If $F$ gives no information deciding whether $\C$ satisfies both $\Omega_1\Omega_2$ {\em vs.} only $\Omega_2$, then the constrained likelihoods, $\Pr{F|\Omega I}$, should be equal.
Because the principle of indifference gives $\Pr{\C|\Omega I} = \frac{1}{|\Omega|}$, and we have shown that $\Pr{\Omega|I} = \text{const.}\times |\Omega|$, it is possible to not only compare energetic hypotheses, such as $F$ {\em vs.} $G$, but also constraints on phase space.
Eq.~\ref{eq:constr} also shows that once information of this type ($\Omega$) has been moved to the right-hand side, then we will not be able to eliminate it using Eq.~\ref{eq:direct}.  Rather, once $\Omega$ has been assumed, then subsequent addition of information will have to include $\Omega$ as part of $\C$ on the right-hand side of Eq.~\ref{eq:LR}. To remove this information and get $\Pr{F|I}$ would require $\Pr{\Omega|F I}$.

\begin{figure}[htbp]
\begin{center}
\includegraphics[width=0.6\textwidth]{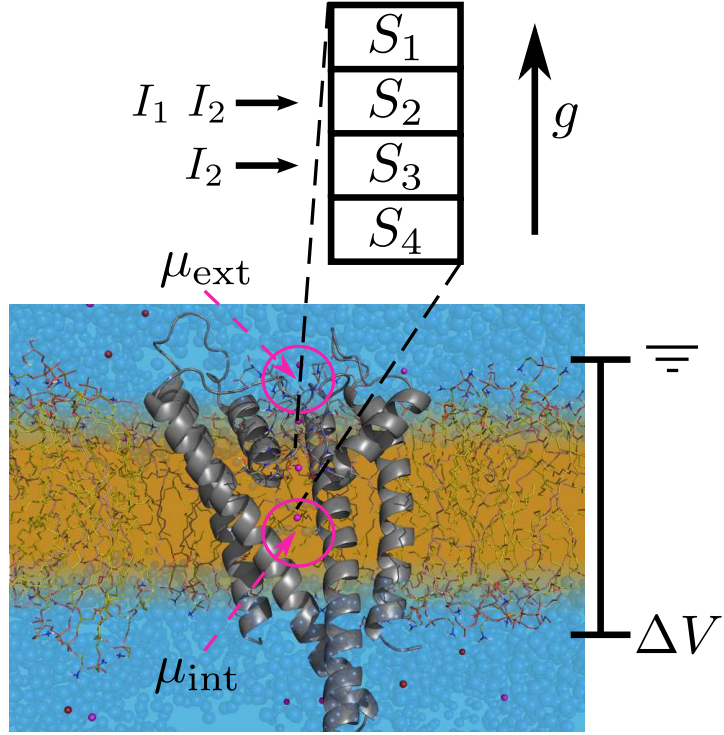}
\caption{KcsA ion channel selectivity filter in its biological orientation (intracellular solution below) showing ion binding sites S1-S4.  For visual clarity, two of the four identical monomer units are not shown.  Physiological conventions for the potential difference, $\Delta V$, and direction of outward positive current ($g$) are indicated.}
\label{fig:filt}
\end{center}
\end{figure}

  As an example, we consider the multi-ion binding site at a K$^+$-ion channel selectivity filter (Fig.~\ref{fig:filt})\cite{jaqvi00}.  Four cationic binding sites are distinguished, and it is assumed that the channel presents a high enough energetic penalty to exclude the possibility of anion occupancy.  We do not expect multiple ion occupancy of the same site to be possible (or highly probable) because of mutual electrostatic repulsion and geometric features of the channel.  This leads us to the fermion-like default statistics,
\begin{equation}
\label{eq:Sx}
\begin{split}
\Sx = \oplus[N_0,N_1\cdot\oplus(X_1,X_2,X_3,X_4),N_2\cdot\oplus(X_1X_2,X_1X_3,X_1X_4,X_2X_3,X_2X_4,X_3X_4), \\
  N_3\cdot\oplus(X_2X_3X_4,X_1X_3X_4,X_1X_2X_4,X_1X_2X_3),N_4X_1X_2X_3X_4]
\end{split}
,
\end{equation}
where $n$ particles may occupy $k$ states in $\binom{k}{n}$ ways for a total of $2^k$ elementary states of the system.  In the absence of any other information, each state is equally likely.
\begin{equation}
\Pr{NX|\Sx I} = \frac{I(NX\in\Sx)}{2^4}
\end{equation}
This probability distribution factors into a product of independent distributions for each site, with equal probability for occupied and unoccupied states.  The distribution is shown for reference in Fig.~\ref{fig:distn}a.

  The partition function is the number of states, $Z[\Sx]=2^4$ (\ref{eq:dZconstr}).  Using the same equation, the partition function of a constrained system, for example at fixed $N$, is $Z[N\Sx]=Z[\Sx]\sum_{X|N} \Pr{NX|\Sx I} = \binom{k}{n}$.  The much debated `degeneracy factor' for particle counting has already crept in as a consequence of the definition (\ref{eq:Sx}), since in the limit $K>>N$, $\binom{k}{n}\to k^n/n!$.
In the following discussion we will successively incorporate mechanical information including the average system energy, and mutual interactions between the ions and the channel.

\subsection{ Addition of Maximum Entropy-Type Information}
\label{sec:maxent}

  The significance of $-\ln Z$ as the (non-dimensional) Gibbs free energy should be immediately recognized.  To show this formally, define a hypothesis, $F$, as
\begin{quote}
$F$: The probability distribution of the system, given that $F$ is accepted, is the most likely observational distribution that obeys $\avg{f(x)|F A}=F$ for any $A$.
\end{quote}
Rather than specifying an absolute state, this information is phrased in terms of the change in the probability distribution from an initial state before $F$ has been accepted.  Representing this prior state of knowledge by $A$, then according to the argument above, the least surprising distribution given information $\avg{f(x)|FA}$ is the maximum entropy distribution.
This distribution should satisfy the mathematical condition,
\begin{equation}
\label{eq:Fent}
 \Pr{X|FA} = \text{argmax } \H[FA|A]\text{  s.t. } \avg{f(x)|FA}=F
.
\end{equation}
The unique solution to this condition is\cite{ejayn03}
\begin{equation}
\label{eq:Fadd}
 \Pr{X|FA} = \frac{\Pr{X|A} \tfrac{dx_A}{dx_{FA}} e^{-\lambda f(x)}}
     {\int \Pr{X|A} \tfrac{dx_A}{dx_{FA}} e^{-\lambda f(x)} \; dx_{FA}}
,
\end{equation}
for some $\lambda(A)$, proving that the hypothesis $F$ (Eq.~\ref{eq:Fent}) is logically equivalent to assuming the probability assignment of Eq.~\ref{eq:Fadd}.  The Jacobian $\tfrac{dx_A}{dx_{FA}}$ has been explicitly shown in this equation because of the importance of continuous functions in thermodynamics.  In a discrete setting, it has the effect of dividing $\Pr{X|A}$ to maintain its normalization.
At this solution, the value of $\H$ is
\begin{equation}
\label{eq:Hadd}
\H_\text{max}[FA|A] = \lambda F + \ln\int \Pr{X|A} \tfrac{dx_A}{dx_{FA}} e^{-\lambda f(x)} \; dx_{FA}
\end{equation}

  According to Bayes' theorem,
\begin{equation}
\label{eq:Fbayes}
 \Pr{F|XA} = \frac{\Pr{X|F A}dx_{FA} \Pr{F|A}}{\Pr{X|A} dx_A}
  = \frac{\Pr{F|A} e^{-\lambda f(x)}}
         {\int \Pr{X|A} \tfrac{dx_A}{dx_{FA}} e^{-\lambda f(x)} \; dx_{FA}}
\end{equation}

  To find the probability of $F$ from a given $X$, we consider two cases.  First, assume $X$ (and $I$) constitute the only data relevant to deciding the plausibility of $F$.  Then $\Pr{F|XAI}=\Pr{F|XI}$ and the terms involving $A$ must evaluate to a constant in the above, so that
\begin{equation}
\Pr{F|XAI} = \text{const}(I)\; e^{-\lambda f(x)} \;\;\text{case 1}
.
\end{equation}
According to the principle of indifference, the leading constant must not depend on $F$, and thus is present to remind us that we are only able to compute likelihood ratios.  We thus have the Boltzmann weight
\begin{equation}
\label{eq:Fwt}
w_F(XA) = \frac{\Pr{F|XA}}{\Pr{\Phi|A}} = e^{-\lambda f(x)}
.
\end{equation}

  In the second case, we may split $A$ into two pieces of information, $B$, determining some weighting over a set of hypotheses of which $F$ is a member, and other information, $A'|B$, irrelevant to $F$ when $X$ is known.  Obviously, maximum-relative entropy hypotheses fall into $A'$, since they are making statements about $X$ and not other hypotheses.  Therefore,
\begin{equation}
\Pr{F|XA'BI} = \text{const}(F;B)\; e^{-\lambda f(x)}  \;\;\text{case 2}
.
\end{equation}
The information, $B$, thus functions as a nuisance parameter\cite{ejayn03} because different assumptions lead to different assignments of plausibilities among $F$ among some class that $B$ affects, and we have
$w_F(XA) = L(F;A)e^{-\lambda f(x)}$.  Because this type of information leads naturally to consideration of alternate classes of hypotheses, we recognize this dividing information to be associated with a set, $\Omega$, of hypothesis space.  If $B$ re-weights relative likelihoods among alternate $F\in\Omega$, then $F$ has effectively become a coordinate and $B$ an energy-type constraint.  If $B$ re-weights all $F\in\Omega$ by the same amount, then its effect is to shift $\Pr{\Omega}$.
We therefore arrive at the diagram picture of Fig.~\ref{fig:rxn}.  Relative likelihoods between nodes can be computed via Eq.~\ref{eq:direct} (case 1) or~\ref{eq:LR} (case 2).  Subgraphs of this structure represent thermodynamic cycles.

  Sequentially using the maximum-relative entropy hypothesis, $F$, requires special consideration of the order in which information is added.  For this type of constraint, the probability distribution is found to be {\em independent} of the order of information addition.  This can be verified by recursion, writing the result of applying Eq.~\ref{eq:Fadd} twice.  Surprisingly, the relative entropies add to the state function Eq.~\ref{eq:Hstate}.  Starting from $\Sx$ and moving to $F \Sx$ gives
\begin{equation*}
\H[F\Sx|\Sx] = \sum_{X\in\Sx} \Pr{X|F\Sx} \ln \frac{\Pr{X|\Sx}}{\Pr{X|F\Sx}}
\end{equation*}
Adding $F\Sx\to FG\Sx$ gives
\begin{align*}
\H[FG\Sx|F\Sx] &= \sum_{X\in\Sx} \Pr{X|FG\Sx} \ln \frac{\Pr{X|F\Sx}}{\Pr{X|FG\Sx}} \\
  &= \H[FG\Sx|\Sx] + \sum_{X\in\Sx} \Pr{X|FG\Sx} \ln \frac{\Pr{X|F\Sx}}{\Pr{X|\Sx}} \\
  &= \H[FG\Sx|\Sx] - \lambda F - \ln\frac{\Pr{F|\Sx}}{\Pr{\Phi|\Sx}} \\
  &= \H[FG\Sx|\Sx] - \H_\text{max}[F\Sx|\Sx]
.
\end{align*}
Therefore, when $F$ is a maximum-relative entropy hypotheses,
\begin{equation}
\label{eq:Htrans}
\H_\text{max}[FGA|A] = \H_\text{max}[FGA|FA]+\H_\text{max}[FA|A]
.
\end{equation}

  Jaynes\cite{ejayn03} has used the functional Eq.~\ref{eq:H} and Eq.~\ref{eq:Z} to derive a host of general relations for maximum entropy constraints including the computation of averages,
\begin{equation*}
\avg{f_j(x)|\C} = -\pd{\ln Z[\{F_j\}\Sx]}{\lambda_j}
,
\end{equation*}
and the (Legendre transform of the) first law of thermodynamics
\begin{equation}
\label{eq:one}
d(-\ln Z[\{F_j\}\Sx]) = \sum_j \avg{f_j(x)|\C} d\lambda_j - \pd{\ln Z}{|\Sx|} d|\Sx|
,
\end{equation}
from which the Gibbs relations,
\begin{equation}
\label{eq:Gibbs}
\frac{\partial^2 \ln Z[\C]}{\partial \lambda_i\lambda_j} = \avg{f_i(x)f_j(x)|\C} - \avg{f_i(x)|\C}\avg{f_j(x)|\C} \\
 = -\pd{\avg{f_j(x)|\C}}{\lambda_i} = -\pd{\avg{f_i(x)|\C}}{\lambda_j}
,
\end{equation}
may be found.  We find that it is more appropriate to phrase these relationships in terms of Legendre transforms of the entropy functional, $\F\equiv\sum_j \lambda_j \avg{f_j(x)} - \H$ for the specific problems considered in Sec.~\ref{sec:cond}.  This distinction was unnecessary before because $\F=-\ln Z$ for distributions derived strictly from constraints of the maximum-entropy form.

  A central maximum entropy constraint in statistical mechanics is a constraint on average energy.  We label this constraint by $\beta$.  A simplified energy function is constructed for the ion channel system by including a mutual Coulomb repulsion between the ions, constrained to the vertical axis and spaced at 3.5\AA{}.  We also assume a simple stabilization energy for each ion from the protein, $E^0 \approx -115$ kcal/mol.
Abbreviating $NX$ to $X$, the energy function is
\begin{equation}
E(X) = E(n,x) = \frac{1}{2} \sum_{i\ne j} \frac{q^2}{4\pi\epsilon_0 |x_i-x_j|} I(X_iX_j) + \sum_i E^0 I(X_i)
.
\end{equation}
Placing this constraint on the average system energy at constant $N$ leads to the well-known canonical distribution with partition function
\begin{align*}
Z[N\beta\Sx] &= \frac{\Pr{N\beta\Sx|I}}{\Pr{\Phi|I}\Pr{\varphi|I}} \\
&= Z[N\Sx] \sum_{X} \Pr{X|N\Sx} w_\beta(NX) = \binom{4}{n} \sum_{X|N} \binom{4}{n}^{-1} e^{-\beta E(X)}
.
\end{align*}
Here, it can be seen that the probability for $N\Sx$, $\binom{4}{n} \Pr{\varphi|I}$ (\ref{eq:const}), cancels in the expression so that the increment $Z[N\beta\Sx]/Z[N\Sx]$ is an average according to Eq.~\ref{eq:LR}.  Removing the constraint on $N$ also leads to the multicanonical ensemble in the same way, {\em viz.} $Z[\beta\Sx]=\sum_N Z[N\beta\Sx]$ (Eq.~\ref{eq:dZconstr}), $\Pr{N|\beta\Sx}=Z[N\beta\Sx]/Z[\beta\Sx]$.

  In either case, we can assign the parameter $\beta$ the meaning of, ``there exists a physical mechanism that decreases the likelihood of the system being in a high-energy state.''  To separate these energy states, we introduce a constraint on the energy, denoted by $E$.  Thus, if a system were allowed to choose its own energy state\footnote{Alternatively, to avoid anthropomorphic terminology, if the system energy is not constrained and we compare the maximum entropy $\Pr{E|A}$.}, the force would bias this choice according to $\Pr{E\beta|A}/\Pr{E\Phi|A}=e^{-\beta E}$.  We can set this bias, $\beta$, to give a reference system with known properties by exactly balancing its internal tendency toward higher energy, $\Pr{E+dE|A}/\Pr{E|A} e^{-\beta dE} = 1$.  This implies that $\beta$ should solve $\beta=\pd{}{E}\ln Z[EA]$ for a reference system with known energy, for example a thermometer in which energy is easily measured by size expansion.  Because our reference thermometer is constantly exchanging energy with the environment, we usually observe its average energy, and $\beta$ should be chosen such that $\avg{E|\beta A}=-\pd{}{\beta}\ln Z[\beta A]$.  The difference between these values (maximum {\em vs.} average energy) is important for small systems, but becomes negligible in the limit of large system sizes.\cite{callen}  Using either of these forces in the present system mimics the effect of allowing energy exchange between the thermometer at this state and the system.  This explains the convention of identifying temperature with the dilation of a thermometer and its connection to the statical force, $\beta$.

\begin{figure}[htbp]
\begin{center}
\includegraphics[width=0.6\textwidth]{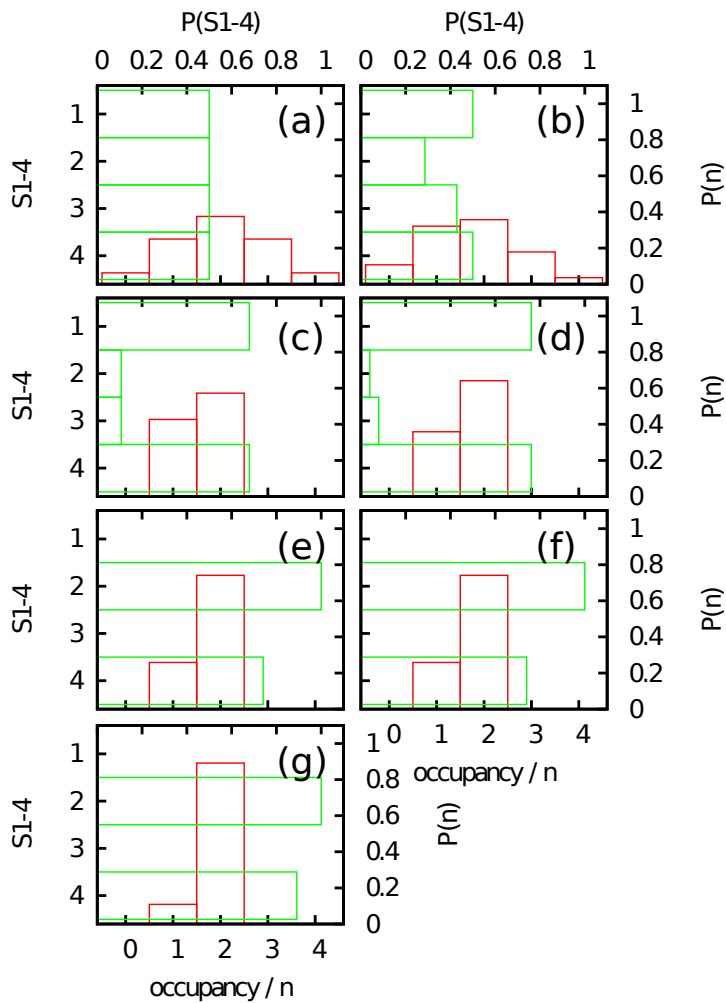}
\caption{Ion occupancy distribution in successively complex models.}
\label{fig:distn}
\end{center}
\end{figure}

  Another constraint we may add is the inclusion of an external force on the total number of ions, $\mu$.  Because the $n$ ions are more likely to choose an environment with lower energy, $-\mu n$, this changes the probability of ion occupancy by $\tfrac{\Pr{\mu|N}}{\Pr{\Phi|N}} = e^{\beta\mu n}$.  The multiplier $\beta$ appears because we want to express $\mu$ in energy units.  Just as above, we can choose the chemical potential, $\mu$, to give a reference system with known properties by balancing its internal energy change on ion addition using the choice $(\beta\mu)=-\ln \frac{Z[(N+1)A]}{Z[NA]}$.\cite{pdt}  We can mimic the effect of allowing K$^+$ transfer from a bulk 100 mM KCl solution to the present system (with the corresponding Cl$^-$ moved to a similar environment and its contribution neglected) by choosing $\mu_\text{K$^+$}=-81+\beta^{-1}\ln 0.1$ kcal/mol.\cite{hfried73}  Without the constraint on $N$, the system was effectively allowed to exchange particles with vacuum.  The combination of both constraints, which we refer to as $F=\beta\mu$, is shown in panel (c) of Fig.~\ref{fig:distn}.  The preference for the separated state ($X_1X_4$) in this model shows the effect of mutual ion repulsion.

\subsection{ Addition of Variables (Generalized Ensemble Methods)}
\label{sec:multi}

  The theoretical background in Sec.~\ref{sec:info} allows us to go further than the most common relations of thermodynamics summarized in the last two subsections.  In particular, the choice of coordinate space, $\Sx$, is no different than any other constraint except that it is almost never moved to the left-hand side to form quantities such as $\Pr{F\Sx|I}$ and comparisons between states are carried out almost exclusively with a fixed $\Sx$.  The addition of coordinates is associated with the transition from canonical to multicanonical ensembles.  It has served as the starting point for some very difficult reading in thermodynamics textbooks involving over/under counting and (in)distinguishability arguments.  Our definition of $\Pr{\Omega|I}$ (\ref{eq:pom}) counts each `state of knowledge' once, and thus directly accounts for (in)distinguishability factors.  As will be shown in a subsequent paper, this result does not require input from quantum mechanics other than a specification of the allowed states of the system.

  Since the rules have already been given above, we proceed to an example, addition of protein-ion interactions by assuming a set of protein conformational states.  This leads to the conception of a generalized ensemble.
A simplistic example is provided by assuming (in addition to an open state, $O$) two `C-type' inactivated states in which a pinching motion of the pore prevents occupancy at site 2 (state $I_1$) or sites 2 and 3 (state $I_2$)\cite{lcuel10}.
These states are assumed to be mutually exclusive and exhaustive, so that all conformational states, $Y$, are a member of the space $\Omega=\oplus(O,I_1,I_2)$.
Before any coupling is assumed, the total number of occupancy states, $|\Sx|$, is multiplied $|\Omega|$ times to create the product space, $\Omega\times\Sx$.  When the conformational state is known, $\Omega$ is irrelevant, and we can intuitively use the knowledge of its coupling to $X$ (denoted by $G$) to guess a form for $\Pr{X|YFG\Sx}$.  This is an instance where intuition runs ahead of logical reasoning, and it is difficult to see the logical steps required to arrive at this result.

  Because $X$ is coupled to $Y$, the conformation, $Y$ is also coupled to the occupancy state, $X$, and it is necessary to know the full distribution, $\Pr{XY|FG\Sx\Omega}$.  This can be rationally arrived at using our thermodynamic diagram (Fig.~\ref{fig:rxn}).  We could first add a non-interacting space for $\Omega$ to get $\Pr{XY|F\Sx\Omega}=\Pr{X|F\Sx}\Pr{Y|\Omega}$ ($F\Sx\to F\Sx\Omega$) and then add information on their coupling, $F\Sx\Omega\to FG\Sx\Omega$.  The distribution of $X$ changes when $G\Omega$ is known, since $G$ places constraints on both $X$ and $Y$.
\begin{equation}
\Pr{X|YFG\Sx} = \frac{\Pr{XG|YF\Sx}}{\Pr{G|YF\Sx}} = \frac{\Pr{XG|YG\Sx}}{\sum_{X\in\Sx} \Pr{XG|YG\Sx}}
\end{equation}

  Is there more to learn from this result?  In Sec.~\ref{sec:info} we showed that any order of adding the information leads to equivalent results, as long as free energy differences are computed in the direction of increasing constraints.  Given information $FG\Omega\Sx$ (or $FGY\Sx$), we are able to write down the distribution for $XY$ simply by maximizing the entropy $\H[FG\Omega\Sx|\Omega\Sx]$ (or $\H[FGY\Sx|\Sx]$).  These generate conditional distributions given information of the type: `the system is in a given coarse state.'  The unanswered question is what the distribution over the coarse states looks like.  To answer this, we consider the process $F\Sx\to FG\Sx\Omega$.  The mutually exclusive and exhaustive condition, $\Omega$, defines a space for the coarse coordinates, $Y$.  However, without this space, we may still calculate $F\Sx\to YFG\Sx$,
\begin{align*}
\frac{Z[FGY\Sx]}{Z[F\Sx]} &= \frac{\Pr{FGY\Sx|I}}{\Pr{\Phi|I}\Pr{F\Sx|I}} \\
 & = \frac{\Pr{GY|F\Sx I}}{\Pr{\Phi|I}} = \sum_{X\in\Sx} \frac{\Pr{GY|FX I}}{\Pr{\Phi|I}} \Pr{X|F\Sx}
.
\end{align*}
This could also have been arrived at through the intermediary path $F\Sx\to YF\Sx\to YFG\Sx$.  The probability for $Y$ in some mutually exclusive and exhaustive set is a sum of these
\begin{align*}
\frac{Z[FG\Omega\Sx]}{Z[F\Sx]} &= \frac{\Pr{FG\Omega\Sx|I}}{\Pr{\Phi|I}\Pr{F\Sx|I}} \\
 & = \sum_{Y\in\Omega} \frac{\Pr{GY|F\Sx I}}{\Pr{\Phi|I}} = \sum_{Y\in\Omega} \frac{Z[FGY\Sx]}{Z[F\Sx]}
.
\end{align*}
We find again that the partition function of Eq.~\ref{eq:Z} has a direct probability interpretation as an un-normalized probability.

  This idea forms the basis for understanding the free energy difference as a log-likelihood ratio between two Hamiltonians  as expressed by Eq.~\ref{eq:DF} and for extending a canonical ensemble into a multi-canonical one.  To perform the extension, define some space over which a previously fixed parameter may vary, and then integrate the partition function over this space.  Given a set of mutually exclusive and exhaustive coarse states, we may write down the micro/multi split using
\begin{align}
\label{eq:multivar}
\Pr{XY|FG\Sx\Omega} &= \Pr{X|YFG\Sx}\Pr{Y|FG\Sx\Omega} \\
\intertext{and the coarse probabilities using either of}
 \Pr{Y|FG\Sx\Omega} &= \frac{\sum_{X\in\Sx} \Pr{XYFG|\Sx\Omega}}{\sum_{Y\in\Omega}\sum_{X\in\Sx} \Pr{XYFG|\Sx\Omega}} \\
\label{eq:multi}
                    &= \frac{Z[YFG\Sx]}{\sum_{Y\in\Omega} Z[YFG\Sx]}
.
\end{align}
The denominators of the second and third expressions correspond to the free energies for processes $\Sx\Omega\to FG\Sx\Omega$, and $\Phi\to FG\Sx\Omega$, respectively.  This argument holds when $Y$ denotes any type of constraint, and the generalized ensemble method is an example of the above when $Y$ are alternate Hamiltonians.\cite{uhans96,elym06}

  As an aside, the interpretation of Eq.~\ref{eq:H} given in the introduction implies that the relative entropy addition $\Sx\to\Sx\Omega$ (as well as $F\Sx\to F\Sx\Omega$) is zero.  This is a reasonable result in the following sense.  If some distribution over $X\in\Sx$ is assumed, and new observations of a coordinate, $Y$, became available that were nevertheless completely random, then $F\Sx\Omega$ does not have any additional informational value, relative to $F\Sx$.  This is contrary to the behavior of the thermodynamic entropy because the thermodynamic entropy increases whenever states are added to the system, even if they are irrelevant, leading to nonzero entropy for nuclear spin systems at zero Kelvin.  Instead of this behavior, it seems preferable to define the entropy relative to the completely uniform distribution, as we have done here.  In this case, the probability for occupying degenerate (but distinguishable) states increases because of the counting conventions of the free energy functional.

  Incorporating the conformational state information, $G\Omega$, into the ion channel system leads to the results shown in panels (b, no energetic constraint) and (d, constrained chemical potential and energy) of Fig.~\ref{fig:distn}.  Because fewer states are available to the system in conformations $I_1$ and $I_2$, they appear less often.  Colloquially, they are said to be entropically un-favorable.  In our derivation, this entropy decrease came about from adding information $G$.  This result that could have been derived either as a consequence of formally reducing the number of occupancy states (as we have done) or by assuming a very large energy for un-allowed occupancies at $I_1$ and $I_2$.  The statement, `$I_2$ is entropically unfavorable' is therefore expressing the fact that the accessible volume for $X$ has decreased from some previously available volume upon changing $\Omega$ to $I_2$ or upon adding information $I_2 G$.  The conventional thermodynamic entropy implicitly defines this previously available volume, regardless of whether such a state physically exists.  This dependence is made explicit in the present definition of a relative entropy.


\subsection{ Conditional Maximum-Entropy Information}
\label{sec:cond}

  If, instead of the energy function assumed for $F$ in the above example, we had assumed some experimentally known probability distribution over $X$, then adding information $G$ becomes qualitatively different.  In order to not interfere with the distribution over $X$, the information $F$ must take priority over any other constraints we may add to the problem.  However, this does not prevent us from coupling $Y$ to $X$ using the conventional maximum-relative entropy hypothesis,
\begin{quote}
$G$: The probability of $XY$, given that $G$ is accepted, is the most likely observational distribution that obeys $\avg{g(y;x)|AXG} = G(X)$ for any $AX$.
\end{quote}

  This is because the entropy functional decomposes as
\begin{align}
\H[AG\Sx\Omega|A\Sx\Omega] &= \sum_{XY} \Pr{XY|AG\Sx\Omega} \ln \frac{\Pr{XY|A\Sx\Omega}}{\Pr{XY|AG\Sx\Omega}} \notag \\
 &= \sum_{XY} \Pr{XY|AG\Sx\Omega} \left[
   \ln \frac{\Pr{X|A\Sx\Omega}}{\Pr{X|AG\Sx\Omega}} +
   \ln \frac{\Pr{Y|AX\Omega}}{\Pr{Y|AGX\Omega}} \right] \notag \\
\label{eq:condent}
 &= \H_X[AG\Sx\Omega|A\Sx] + \sum_X \Pr{X|AG\Sx\Omega} \H_Y[AGX\Omega|AX\Omega]
.
\end{align}
The sums in this section are all taken to be over $X\in\Sx$ and $Y\in\Omega$ without loss of generality since we choose $\Sx\times\Omega$ to be the set of all $XY$ relevant to deciding $A$ or $G$.
The last term in the expansion above is a conditional entropy, which is a functional of $\Pr{Y|AGX\Omega}$ and depends on $X$.  Because each conditional distribution can be chosen independently from the others and from $\Pr{X|AG\Sx\Omega}$, the entropy of each one is independently maximized when $\H[AG\Sx\Omega|A\Sx\Omega]$ is maximum.  However, the presence of $Y$ allows $\H_X[AG\Sx\Omega|A\Sx]$ to differ from $\H_X[A\Sx|A\Sx]=0$, since $\Pr{X|AG\Sx\Omega}=\sum_Y \Pr{XY|AG\Sx\Omega}$.  For these two to be equal in general requires that $\Pr{X|AG\Sx\Omega}=\Pr{X|A\Sx}$ -- i.e. that the distribution of $X$ is not dependent on the information $G\Omega$ when $A$ is present.

  Because we want to specify the marginal distribution of $X$ directly, it is convenient to denote this information as the compound hypothesis,
\begin{quote}
$F_X$: The probability distribution of $X$ is determined by information $F_X$ and unchanged by information $G\Omega$.
\end{quote}
When this hypothesis is in place, we will have $\Pr{X|F_XG\Sx\Omega}=\Pr{X|F_X\Sx}$.  Bayes' theorem says that we must also have $\Pr{G\Omega|XF_X\Sx}=\Pr{G\Omega|F_X\Sx}$, implying $w_{G\Omega}(F_XX)=1$.  Effectively, the $Y$ have become `imaginary states' to the system in the sense that there is no free energy change for $F_X\Sx\to F_XG\Sx\Omega$.

\begin{figure}[htbp]
   \centering
   \includegraphics[width=0.9\textwidth]{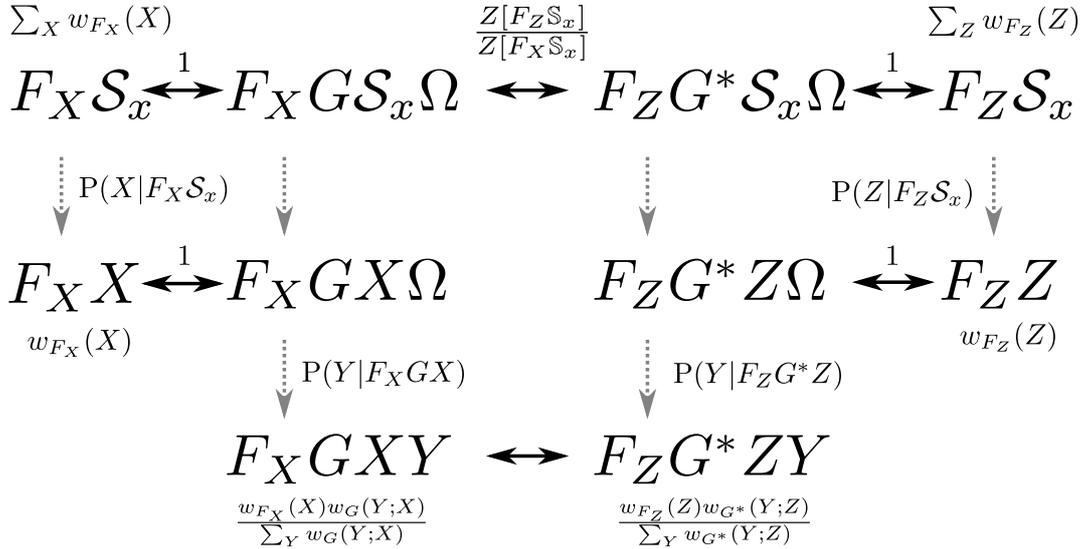} 
   \caption{Reaction diagram for adding conditional maximum entropy information.  Partition functions, determined by likelihood ratios for each transition, are written out for each state.  For the `forward' process $F_X\Sx\to F_XG\Sx\Omega$, there is a `reverse' process $F_Z\Sx\to F_ZG^*\Sx\Omega$ signifying the dual maximum conditional entropy problem.}
   \label{fig:cond}
\end{figure}

  Although there is no change to $\H_X$ or the distribution of $X$, maximizing (\ref{eq:condent}) results in
\begin{equation}
\label{eq:infer}
\Pr{Y|F_XGX\Omega} = \frac{\Pr{Y|F_XX\Omega} e^{-\lambda g(y;x)}}
        {\sum_{Y\in\Omega} \Pr{Y|F_XX\Omega} e^{-\lambda g(y;x)}}
,
\end{equation}
an expression reminiscent to the transition probability for a Markov process.
The conditional entropy is
\begin{align*}
\H_Y[F_XGX\Omega|F_XX\Omega] &= \sum_{Y\in\Omega} \Pr{Y|F_XGX\Omega}\ln\frac{\Pr{Y|F_XX\Omega}}{\Pr{Y|F_XGX\Omega}} \\
 &= \avg{\lambda g(y;x)|F_XGX\Omega} + \ln \sum_{Y\in\Omega} \Pr{Y|F_XX\Omega} e^{-\lambda(x) g(y;x)}
,
\end{align*}
and we define as usual
\begin{equation*}
w_G(XY) = \frac{\Pr{G(X)|XYI}}{\Pr{\Phi|XYI}} = e^{-\lambda g(y;x)}
.
\end{equation*}
These considerations are sufficient to fill out the thermodynamic cycle when $F_X$ is assumed, as has been done in the left half of Fig.~\ref{fig:cond}.

  Imposing the distribution among ion occupancy states given in Ref.~\citenum{jaqvi00} (shown for reference in Fig.~\ref{fig:distn}f) as $F_X$, application of this procedure to determine the conformational equilibrium shows that the channel is almost always in the open state due to the high probability for occupancy of S2.  The probabilities for $I_1$ and $I_2$ are 2.3$\cdot 10^{-4}$ and 8$\cdot 10^{-6}$.
Although $X|F_X$ is independent from $G\Omega$, knowledge of $Y$ is still informative for $X$, as
\begin{equation}
\Pr{X|YF_XG\Sx} = \frac{\Pr{X|F_X\Sx} \Pr{Y|F_XGX\Omega}}{\sum_{X'} \Pr{X'|F_X\Sx} \Pr{Y|F_XGX'\Omega}}
.
\end{equation}
Using this method of inference, the occupancy distribution in the open state is shown in Fig.~\ref{fig:distn}e.  There is a very slight increase in occupancy at S2 and a decrease at S3, but the effect is small because the open structure is dominant.  Note that our assumption that the free energies of Ref.~\citenum{jaqvi00} are averages over the conformational states was chosen for illustration and may be incorrect.

  We argue that addition of this type of conditional information is central to non-equilibrium statistical mechanics.  To derive an ensemble of trajectories, we add all possible transitions, $Y$, originating from each state, $X$.  The initial state and its transitions are linked by some information, $G$, which determines the distribution of $Y$ given $X$.  This constraint determines a maximum entropy {\em transition} probability density, as considered in differential form in Refs.~\citenum{pberg55,jlebo59}.  The hypothesis $F_X$ states that what we know about the starting distribution is completely determined by $F_X$ and not by any possible, but unknown, future events.  It is required for the process to be non-anticipating in the sense that no information about processes we may carry out in the future -- $G\Omega$ is available from $X$. 

  There is a great deal of literature on methods for non-equilibrium statistical mechanics.  Because this paper is intended to show a new way of approaching problems, we will confine ourselves to deriving two main results of the non-equilibrium theory.  The first is the more recent development of fluctuation formulas for irreversible entropy.  Fig.~\ref{fig:cond} displays the duality between fixing $F_X$ at the initial time and fixing its propagated distribution $F_Z$.  In setting up an inference problem for $Y$ starting from $F_X GX$, the distribution of $Y$ is given by (\ref{eq:infer}).  If this distribution is used to determine $F_Z$ using $\Pr{Z|F_XG\Sx}=\sum_{XY} \Pr{Y|F_XGX\Omega} \Pr{X|F_X\Sx} I(Z=Z(Y))$, some information loss occurs when $F_X$ is discarded and only $F_Z$ and information constraining the transitions between states, $G$, retained.  Assuming the transitions, $Y$, specify both end-points $X$,$Z$, the distribution of $Y$ carries the complete information for this process.  Using the information loss metric\cite{hakai71,ejayn03},
\begin{align}
L &= -\H[F_XG\Sx\Omega|F_ZG^*\Sx\Omega] \notag \\
  &= \sum_{Y} \Pr{Y|F_XG\Sx\Omega} \ln \frac{\Pr{Y|F_XG\Sx\Omega}}{\Pr{Y|F_ZG^*\Sx\Omega}} \notag \\
  &= \avg{\ln\frac{\Pr{Y|F_XGX\Omega}}{\Pr{Y|F_ZG^*Z\Omega}} + \ln\frac{\Pr{X|F_X\Sx}}{\Pr{Z|F_Z\Sx}}} \notag \\
\label{eq:loss}
  &= \H_Z[F_Z\Sx|\Sx] - \H_X[F_X\Sx|\Sx] + \avg{\ln \frac{\Pr{Y|F_XGX\Omega}}{\Pr{Y|F_ZG^*Z\Omega}}}
.
\end{align}
The averaging is taken in the forward direction, and so $L\ge 0$ evidently represents the amount by which the real distribution $F_XG\to XYF_XG$ contains information not present in a distribution guessed from $F_ZG^*$.  Note that if $G$ allows only one-to-one $XZ$, the transitions are deterministic, and zero information is lost.  More generally, if forward and backward inference directions yield the same joint distribution so that $F_XG=F_ZG^*$, then there is no way to discern the direction of time's arrow and no information is dissipated.

  The above relations are purely statistical, and have been stated in terms of maximum entropy constraints for forward, $G$, and reverse, $G^*$, inference problems.  They are generally valid for any choice of $G^*$.  In derivations of the fluctuation theorem,\cite{rkawa07} a particular choice of $G^*$ is made corresponding to time-reversed equations of motion.  The statistical perspective expressed here shows that this operation is confined to the choice for $G^*$, and provides a suggestion as to the informational role of time-reversal.  For example, the forward constraints are consistent with the Langevin equation,
\begin{equation*}
\Pr{Z|X} \propto e^{-(\Delta p-F_X)^2/2\sigma^2-(\Delta p-F)v_X\beta/2}
,
\end{equation*}
so that the momentum change ($\Delta p = p_Z-p_X$) is normally distributed about $F-\gamma v$ to yield a Boltzmann distribution.  The correct choice of $G^*$ is given by changing $\beta$ to $-\beta$ in the above equation.  The equation for Brownian motion can be similarly derived by constraining $\Delta x^2$ with $\sigma^{-2}/2$ and $-\Delta x F/2$ with $\beta$.  In both of these equations, the same set of forward transitions are used for $G^*$, but the sign of the Lagrange multipliers constraining the fluxes are reversed.  We can thus intuitively see that reversing the {\em sign of externally applied forces} gives the correct fluctuation theorems using the information loss metric (Eq.~\ref{eq:loss}).  This relation is valid in transient stochastic dynamics, and allows for entropy to increase both by increasing the entropy of the distribution (first part of Eq.~\ref{eq:loss}) and by the presence of irreversible fluxes (last term of Eq.~\ref{eq:loss}).  Such an informational perspective is required for understanding entropy increase for processes which to not have time-reversal symmetry, but nevertheless have well-defined and reproducible behavior.

  Retaining only information about the end-points of a path $\Gamma=X_1X_2\ldots X_N$, from $F_1$ to $F_N$, we denote $\Gamma_i=X_1\cdots X_i$ and $\Gamma^i=X_i\cdots X_N$.  We also assume constant $\Sx$ and conditional independence, $\Pr{X_{i+1}|G\Gamma_iF_1}=\Pr{X_{i+1}|G\Gamma_i}$.  If the transitions are known from $\Gamma$, the total dissipation is
\begin{equation}
\label{eq:dS}
dS/k_B \equiv L = \H_N[F_N\Sx|\Sx]-\H_1[F_1\Sx|\Sx] + \avg{\sum_{i=1}^{N-1} \ln\frac{\Pr{X_{i+1}|G\Gamma_i\Sx}}{\Pr{X_i|G^*\Gamma^{i+1}\Sx}}}
,
\end{equation}
where $k_B$ is the Boltzmann constant.
This path functional is in agreement with the thermodynamic entropy production given by the ratios of forward and reverse path probabilities\cite{gcroo99,cjarz06,rkawa07} as well as an expression for entropy production deduced from mechanical considerations\cite{pberg55} when $\ln \Pr{X_{i+1}|G\Gamma_i\Sx}/\Pr{X_i|G\Gamma^{i+1}\Sx} = -\lambda g(x_{i+1},x_i)$, with $g$ a generalized flux.  We have derived this result from the direction of information propagation,\cite{ejayn57a} and no special treatment has been given to the multiplier, $\beta$, defining the externally applied temperature.  This derivation also avoids the complications associated with defining a steady-state.  A curious feature is that it does not make specific reference to heat.  This may be explained by noting that the transitions associated to fluxes, $g$, are probabilistic and represent interaction with an external system.  These transitions may add or remove energy from our system, while the external system remains at a fixed thermostatic temperature state, $\beta_\text{ext}^{-1}$.  We then define the heat injected from the environment as the net energy gain, $\beta_{ext} dQ=\avg{\lambda g(x_{i+1};x_i)}$.  This identifies (\ref{eq:dS}) with the Clausius form for the second law,\cite{jvonn96,ejayn84,ejayn92}
\begin{equation}
\label{eq:two}
dS/k_B = dS_\text{int}/k_B - \beta_\text{ext} dQ \ge 0
.
\end{equation}
The above claims relating transition probabilities to fluxes can be established for the Langevin and Brownian equations, and have been more thoroughly explored in a manuscript devoted to nonequilibrium problems\cite{droge11b}.

  The next result will be a derivation of the fluctuation-dissipation theorems from the Gibbs relations.  Because our free energy for the process $A=F_{X1} \mathbb S_{X1} G_{12} \mathbb S_{X2} G_{123} \ldots$ is simply the free energy for $F_{X1} \mathbb S_{X1}$, we must find an alternate free energy functional.  The `caliber' function of Jaynes,
\begin{equation}
\H[A\mathbb S_\Gamma|\mathbb S_\Gamma]
  = \sum_\Gamma \Pr{\Gamma|A\mathbb S_\Gamma} \ln
          \frac{\Pr{\Gamma| \mathbb S_\Gamma}}
               {\Pr{\Gamma|A\mathbb S_\Gamma}}
\end{equation}
lends itself to the task by defining the Legendre transform
\begin{align}
\F[\lambda] &\equiv \sum_{i=1}^N \avg{\lambda_i g_i(\Gamma_i)|A\mathbb S_\Gamma} - \H[A\mathbb S_\Gamma|\mathbb S_\Gamma] \notag \\
 &= -\sum_\Gamma \Pr{\Gamma|A\mathbb S_\Gamma} \ln \prod_{i=1}^N \avg{e^{-\lambda_i g_i(X_i;\Gamma_{i-1})}\Big|\Gamma_{i-1}} \notag \\
\label{eq:nonF}
 &= \sum_{i=1}^N \avg{-\ln Z[G_i\Gamma_{i-1}\Sx]\;\big|A\mathbb S_\Gamma}
.
\end{align}

  The first derivatives generate a `first law' for non-equilibrium processes,
\begin{align}
\pd{\F}{\lambda_i} &= \avg{g_i(\Gamma_i)} \notag \\
d\F &= \sum_i \avg{g_i(\Gamma_i)} d\lambda_i
.
\end{align}
This is a path average conditional on $A\mathbb S_\Gamma$, but this notation has been suppressed for clarity.
The second derivatives are the Green-Kubo formulae
\begin{align}
\label{eq:GK}
\frac{\partial^2 \F}{\partial \lambda_i \partial \lambda_j} = -\avg{\delta g_i(\Gamma_i)\delta g_j(\Gamma_j)}
\end{align}

  For the ion channel example we have been developing, a completely new set of constraints must be developed for transitions between states.  
For the forward problem, we are given $X_i$ as well as some set of feasible transitions, $Y|X_i$, from state $i$.  Because the probability of inactivated states are negligible, we consider only the open channel state, and single-jump transitions as shown in Fig.~2 of Ref.~\citenum{jaqvi00}.  Five transitions from each state are possible, corresponding to doing nothing, or all sites moving up or down by the addition of a K$^+$ or a water at the appropriate end.

  In order to produce a system that conserves energy, we place a constraint on the energy change at each step.
\begin{equation}
\Pr{Y|X_i\beta'\Omega} = \frac{e^{-\beta (E(X_{i+1})-E(X_i))}}{\sum_{Y|X_i} e^{-\beta (E(X_{i+1}(Y))-E(X_i))}}
\end{equation}
This amounts to a stochastic addition of energy to the system in the amount of $\avg{dE|X_i}=-\pd{Z[X_i\beta'\Omega]}{\beta'}$.  The steady-state distribution will differ from the canonical distribution in general because the normalization constant, $Z[X_i\beta'\Omega]$, depends on $X_i$.  This difference has come about because of the addition of information limiting which transitions are possible.  If all states were available during each transition, the normalization constant would again be independent of $X_i$ and we would recover the canonical distribution.  For the Langevin and Brownian equations with uniform applied temperature, the canonical distribution is also obtained because the normalization constant is independent of $X_i$.

  Because transitions are not generally spontaneous, but may have an energy barrier, we add another constraint, $\beta'E^\dagger$, directly on the number of transitions per time-step, $\tau$,
\begin{equation}
\Pr{Y|X_i\beta'E^\dagger\Omega} = \frac{\Pr{Y|X_i\beta'\Omega} e^{-\beta E^\dagger I(Y)/\tau}}{\sum_{Y|X_i} \Pr{Y|X_i\beta'\Omega} e^{-\beta E^\dagger I(Y)/\tau}}
.
\end{equation}
These barriers could, of course, be made to depend arbitrarily on the transition, $Y$, but for simplicity we assume that they are present only when a transition occurs and uniformly equal to the sum of $2$ ps kcal/mol.  The stochastic process specified by these two formulas has the identity matrix as the small time-step limit, and an equilibrium-like distribution as the large step limit.  The energy barrier assumption differs from the usual rate equation formulation, since the Chapman-Komologrov equation no longer holds.  Instead, the behavior of the above system is dependent on the time-scale studied, reminiscent of fractal kinetic models.\cite{llieb87}  Note also that $E^\dagger$ may be a function of the time-step, $\tau$, to give a specified average number of transitions to recover a Markov model.  Because this is a novel kinetic model, it remains to be seen how well these two constraints reproduce actual dynamics; however the form of this equation matches well the nonlinearity near $t=0$ in exact transition probabilities computed for the M\"{u}ller-Brown potential surface (Fig.~4 of Ref.~\citenum{dzuck99}), while variations in the surface chosen to divide states can be mimicked by changes in $E^\dagger$.

  To finish our specification of non-equilibrium jump processes, we specify the forces on spontaneous ion creation and annihilation.  Removing the possibility of a change in ion number unless it either enters or exits through an end of the channel, we can then specify the external force, $\mu$, acting on these special events using the same type of energy constraint (and assuming for simplicity the same energy barrier) as above.  This leads to
\begin{equation}
\Pr{Y|X_iA\mu} = \frac{\Pr{Y|X_iA} e^{\beta \mu_\text{int}dN_\text{int}(Y) + \beta \mu_\text{ext}dN_\text{ext}(Y)}}{\sum_{Y|X_i} \Pr{Y|X_iA} e^{\beta \mu_\text{int}dN_\text{int} + \beta \mu_\text{ext}dN_\text{ext}}}
,
\end{equation}
with $dN_\text{int}$ and $dN_\text{ext}$ representing the number of ions added to the system ($\pm 1$) from the internal and external solutions, respectively.  The form of this transition probability is similar to that of a recent paper on currents in boundary driven Kawasaki dynamics,\cite{mbaie09} which were also analyzed using a cumulant-generating function similar to Eq.~\ref{eq:nonF}.

  An outward-driving voltage can be added to the system by imposing an external field, increasing the likelihood for transitions moving ions outward by an amount $e^{\beta \Delta V g(Y)}$.  The function $g(Y)=\sum_j I(X_j\leftarrow X_{j+1})$ counts the number of ions taking a step outward during transition $Y$, consistent with the sign convention of Fig.~\ref{fig:filt}.  For ion movements internal to the channel, this has an equivalent effect on the path distribution as applying an energy constraint $e^{\beta \sum_j V_j I(X_j)}$ ($I(\cdot)$ is the indicator function).  These constraints provide a complete kinetic model for our ion channel in arbitrary solution conditions and driving voltages.
  
  The steady-state ion occupancies at zero applied voltage and $\mu$ identical to that for (e) and (f) of Fig.~\ref{fig:distn} are plotted in panel (g).  The steady-state distribution is slightly altered from the local equilibrium prediction of (e).  This happens despite the fact that the transition probability obeys detailed balance with respect to the steady-state, and exactly five transitions lead into each ion occupancy state.  The reason is that the transition probability is normalized by a different value for the forward and reverse transitions.

  As a final note, the current can be calculated as a perturbation from a steady-state using Eq.~\ref{eq:GK}
\begin{align}
\avg{g(t)}_{\beta\Delta V'} &= -\eval{\pd{\F[\beta\Delta V]}{\beta\Delta V(t)}}{\beta\Delta V'} \notag \\
  &\approx -\eval{\pd{\F[\beta\Delta V]}{\beta\Delta V(t)}}{\beta\Delta V}
  -\sum_{i\le t} \frac{\partial^2 \F}{\partial \beta\Delta V(t) \beta\partial \Delta V(i)}\beta(\Delta V'(i)-\Delta V(i)) \notag \\
  \label{eq:FDT}
  &= \avg{g(t)}_{\Delta V} + \beta \sum_{i\le t} \avg{\delta g_i \delta g_t}_{\Delta V} (\Delta V'(i)-\Delta V(i))
.
\end{align}
This gives the time-dependent linear response for small changes in the holding potential.  The conductance near the resting potential is the time-integral of the steady-state current auto-correlation function (at zero average current), in accordance with Onsager's phenomenological equation\cite{lonsa31}.  The negative sign comes about because of the positive sign of the constraint ($\beta\Delta V$).  At other voltages, this integral is the slope of the current/voltage curve.  The presence of an additive constant explains why Onsager reciprocity only holds near equilibrium, where the fluxes are zero.  Other Legendre transforms of Eq.~\ref{eq:nonF} lead to relationships at fixed current, etc. as in the usual theory.\cite{callen}

\begin{figure}[htbp]
   \centering
   \includegraphics[angle=-90,width=0.5\textwidth]{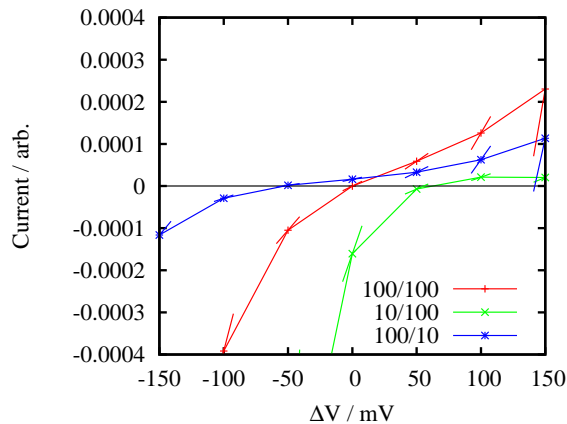} 
   \caption{Current-voltage plot calculated using the free energies from Fig.~2 of Ref.~\citenum{jaqvi00} along with the assumptions listed in the text.  The voltage plotted in this figure is the sum of the five voltage steps between S0-S5.  The integrated autocorrelation function is shown using tangent lines according to the FDT (Eq.~\ref{eq:FDT}).  Although the reversal potential shifts are physically reasonable, inward rectification (opposite known channel behavior) is observed.}
   \label{fig:IV}
\end{figure}

    The current-voltage characteristics calculated for this system are shown in Fig.~\ref{fig:IV}.  The fluctuation-dissipation theorem (Eq.~\ref{eq:FDT}) gives the slope of the current-voltage curve, and is plotted as a tangent line at each data point.  Noticeable deviations occur at positive voltages due to numerical error in calculating the steady-state flux and long-timescale behavior of the autocorrelation function.  This has been traced to very long relaxation times (O(10$^5$) steps) for the current, which is in turn due to the low transition probabilities between conduction states with high free energy barriers.  The set of energy barriers used leads to larger current magnitudes at hyperpolarized voltages (inward-rectifying behavior), inconsistent with the known operation of the channel.  It is of interest to more accurately model the transition energy barriers and determine whether the time-dependence of dwell times for individual states is adequately represented by equations of the present, maximum entropy, form.

\section{ Conclusions}

  This work has attempted to formulate the purely statistical content of statistical mechanics in terms of the Bayesian probability theory of Jaynes\cite{ejayn03}.  From this perspective, thermodynamics is a tool for understanding experimental information and its consequences.  In the process, it has become clear that the principles and mathematical methods are of much more general applicability than conventional arguments would lead one to suppose\cite{ejayn57} and that a large number of advanced concepts and methods can be synthesized in this way.

  Entropy has been defined from the perspective of information theory, representing the (negative) information content of distributions.  Because the entropy is maximized upon adding average value information, its first derivative with respect to variations in the distribution is zero.  The first law of thermodynamics expressed in Eq.~\ref{eq:one} is a direct consequence of this observation.  We have shown that the process of adding average value information while maximizing the relative information entropy at each step is transitive.  Therefore, adding a series of such constraints in any order will lead to the same distribution, with the sum of the information increments adding to the same value for all paths.  Because the entropy was defined only relative to a reference distribution, the information increments are zero whenever the distribution is unchanged by maximizing entropy.  Had this relative form been used to define the thermodynamic entropy, the zeroth law of thermodynamics would not require special treatment of nuclear spin multiplicity at zero Kelvin.

  Thermostatic partition functions, $Z[A\Sx]$, have likewise been identified as expressing relative probabilities.  Changes in this function correspond to changes in information, and can be understood as a subjective probability assignment determining relative likelihoods between allowed alternative states of the system.  Before specifying a set of alternate constraints ($\Omega$) the system may choose between to reach statistical equilibrium, the partition function can only take this relative form, as in ($F\to G$) of Fig.~\ref{fig:rxn}.  Once a complete set of constraints is specified, then the partition function decides the relative probability of each state within $\Omega$, and it is possible to say (Eq.~\ref{eq:multi}) that the probability of state $A$, divided by the probability of $\Omega$, is the probability of $A$ given that $A$ is in the set $\Omega$.  This interpretation of the partition function leads naturally to multicanonical ensemble and umbrella sampling methods\cite{fe07}.

  Comparisons between states of knowledge can be done using these functions, and the picture presented here does not require the specification of a complete set of all possible states of knowledge.  Instead, the relations of Sec.~\ref{sec:info} give a basic, consistent set of equations for defining the changes between these states.  This set already justifies the appearance of (in)distinguishability factors in the partition function, as shown in \S~\ref{sec:constr}.  We have provided a justification for the common indicator function, $w_\Omega(\C)$ (\ref{eq:wom}), for comparing purely entropic changes in phase space, as well as the Boltzmann factor (\ref{eq:Fwt}), for comparing changes in maximum entropy information $\Pr{F|\Sx}/\Pr{G|\Sx} = Z[F\Sx]/Z[G\Sx]$.  We have also shown two more advanced examples, generating the multicanonical ensemble in \S~\ref{sec:multi} and a conditional maximum entropy in \S~\ref{sec:cond}.  These are related to the first two examples as marginal distributions are related to conditional ones.
  
  The concept of building up thermodynamic equations of state by adding system information is important for developing multi-scale understanding of large physical systems.  Because this approach is based on using well-defined system states at each step, the predictions of the coarse-grained theory may be compared with a fully atomistic (or ab-initio electronic) molecular dynamics simulation or coarse-grained Monte-Carlo sampling.  At such levels, the number of states will be greatly increased to include coordinates and momenta of all particles, with a change in the energy function to a more accurate approximation.  Because this level of description quickly becomes computationally intractable, the approximate potential of mean force derived from high-level considerations may be useful for locating important states for detailed study, deriving stochastic boundary conditions, and applying force or energy biasing sampling techniques.

  As is now well known, the statistical machinery outlined here is generally applicable to problems where there is uncertainty.  It can be used equally well in reasoning about equilibrium and coarse-graining problems as well as non-equilibrium processes.  Starting with a `trajectory space' and adding information on allowed transitions as well as expectation values of fluxes between states leads to a state of knowledge about the process.  In such a process, the ability to directly write down the equilibrium distribution (a long-sought goal\cite{wrobi90,rnive09}) disappears in the same way a marginal distribution over coarse-grained variables cannot be directly produced from an equilibrium distribution over all atomistic coordinates and momenta.  Instead, the transition distribution can be directly written, and the transient fluxes and eventual steady-state (if it exists) become path averages.  A consideration of the information loss for stochastic processes leads to a formula similar to the second law of thermodynamics (\ref{eq:dS}), applicable arbitrarily far from equilibrium.  The information entropy functional of the path probability given in Sec.~\ref{sec:info} takes on the definition Jaynes' `caliber,'\cite{ejayn80} while its Legendre transform (\ref{eq:nonF}) is a path free energy functional whose Gibbs relations easily generate Green-Kubo type fluctuation-dissipation theorems.\cite{ejayn79,ejayn80,cmaes99,rdewa03}  We emphasize that these formulas are not required to be extensive or local,\cite{djou88,djou99,skjel08} avoid the necessity of defining a steady-state,\cite{gcroo00,etrep04} and are independent of how we define fluxes so that we do not have to immediately write down hydrodynamic equations.\cite{rluzz02}  The present work has given a necessary statistical foundation for extending these results by carrying over modern equilibrium techniques such as the evaluation of free energy differences\cite{mshir08}, and coordinate/path re-weighting techniques\cite{fytre04,dminh09}.
These formulas achieve Jaynes' goal of providing a ``foundation for the predictive aspect of statistical mechanics, in which a single basic principle and method applies to all cases, equilibrium or otherwise."\cite{ejayn57a}  They imbue non-equilibrium and transient dynamic problems with the same structure as the equilibrium thermodynamics given by Gibbs\cite{jgibb02}, and open the door for a new understanding of processes far from equilibrium.


\begin{acknowledgements}
  This work was supported, in part, by Sandia's LDRD program,
and, in part, by the National Institutes of Health through the NIH Road Map for
Medical Research.
Sandia National Laboratories is a multi-program
laboratory operated by Sandia Corporation, a wholly owned subsidiary
of Lockheed Martin Corporation, for the U.S. Department of Energy's
National Nuclear Security Administration under contract DE-AC04-94AL85000.
\end{acknowledgements}

\appendix
\section{ Formal Derivation of Ratios for Undefined Quantities of Probability}
  There are several axiomatic foundations for probability theory, and the system of Komologrov is perhaps the most widely taught and well-known.  This system begins by assuming a space of elementary states, $\mathbb S_y = \{Y_1,Y_2,\ldots\}$.  We assume that $Y_i$ are mutually exclusive and exhaustive.  In this case, a probability distribution function can be defined as a measure of a set $\Pr{X|\mathbb S_y} = \sum_{Y_i\in X} \Pr{Y_i|\mathbb S_y}$, with $\Pr{\mathbb S_y|\mathbb S_y}=1$ and $\Pr{Y_i|\mathbb S_y} \ge 0 \forall Y_i\in\mathbb S_y$.  Any possible subset of $\mathbb S_y$ then defines an aggregate `state.'  Because they are mutually exclusive, separate elementary hypotheses cannot be combined using the `and' operation, only states, such that the probability of state one and state two is the probability of their intersection, $X_1\cap X_2$.  We have not yet made clear how this structure is related to logical inference.

  Formal logic is concerned with proving logical statements from given assumptions.  Both the assumptions and the statements to be proven can be stated in the form of logical sentences.  Each sentence makes assertions about elementary hypotheses using some combination of the logical operators.  For this paper, we assume the Boolean algebra, including the `or' operation (+), as in $X_1+X_2\equiv$ `$X_1$ or $X_2$', the `and' operation (*), as in $X_1X_2 \equiv$ $X_1$ and $X_2$, and the negation, $\bar X_1 = \text{ not } X_1$.  Logical statements can be assigned a probability by defining states, $X$, as hypotheses of the form `the system is in state $X$.'  Each logical sentence then maps to a set of states by replacing union for `or', intersection for `and', and set complementation for `not.'  Assuming all statements are either true or false, aggregate operations may be defined from these, for example the mutually exclusive statement $A\oplus B=A\bar B+\bar A B$ and the implication $A\Rightarrow B = \bar A+B$.  The probability of a given statement can then be determined from the probability of the set it implies.  In order to make logical inferences from an assumed logical sentence, $\Gamma$, we require the definition of a conditional probability.  This is found by changing the space $\mathbb S_y$ to $\mathbb S_\Gamma$ {\em via} setting $\Pr{Y_i|\Gamma\mathbb S_y}=0 \forall Y_i\notin\mathbb S_\Gamma$ and then re-normalizing, resulting in
\begin{equation}
\Pr{X|\Gamma\mathbb S_y} = \frac{\Pr{X\cap\mathbb S_\Gamma|\mathbb S_y}}{\Pr{\mathbb S_\Gamma|\mathbb S_y}}
.
\end{equation}

  Although very easy to present, this system is unnecessarily restrictive for two reasons.  The first is that it requires the definition of a complete space, $\mathbb S_y$, at the outset, which no means of reasoning can remove to add new hypotheses.  As we have seen, inference can then only take place by successively reducing this space to smaller regions.  By analogy, the process of statistical mechanics would therefore have to begin by assuming a multicanonical ensemble along with its mutually exclusive and exhaustive coordinates, and then derive successively constrained systems.  Although a valid derivation can be produced this way, it appears to deny us the ability to define an isolated physical system.  The second reason is related to this point.  Physically, we would like to begin with the idea of an isolated system and then successively build in more complexity as relevant dynamic variables are discovered.  The Komologrov system does not provide a means of reasoning about a hypothesis without first defining its `space.'  Instead of adding prior information to the right of the conditional sign, we would like to build up a complete picture of a physical system by successively moving prior information over to the left.
  
  This point was considered by Jaynes\cite{ejayn03}, who showed that a probability theory `without bounds' could be derived from three desiderata for assigning plausibilities to logical statements.  The first was that degrees of plausibility be represented by positive, real numbers.  The second and third require that all available prior information is used and that equivalent states of knowledge and reasoning processes lead to identical results.  From these desiderata, the product rule (Bayes' theorem, $\Pr{AB|C} = \Pr{A|BC}\Pr{B|C} = \Pr{B|AC}\Pr{A|C}$), may be deduced.  Here, the only requirement is that A,B, and C represent information and that the postulate, C, does contradict itself.  It is therefore unnecessary to define a space in which $A$ must exist in order to determine its plausibility from $C$.

  One point is worth noting.  A logical sentence of the form $\Gamma=A(B+C)(A\oplus D)$ immediately implies $A$ as well as denies $D$, and provides some information about the statements $B$ and $C$.  However, it does not contain any information whatsoever about an unrelated proposition, $F$.  With some thought, it can be seen that assignment of plausibilities based on a logical sentence must fall into one of four classes: true ($\Pr{A|\Gamma}=1$), possible ($\Pr{B|\Gamma}$), undecidable ($\Pr{F|\Gamma}$), or impossible ($\Pr{D|\Gamma}=0$).  Probability theory is chiefly concerned with propositions that are possible.  However, the product rule also applies to situations in which a proposition is undecidable.

  What has been said serves to illustrate the difficulty of reasoning without assuming a set of mutually exclusive and exhaustive alternatives.  To demonstrate this concretely, we will attempt to assign probabilities to a general logical statement, $\psi$, assuming only the principle of indifference, $I$, and possibly another logical statement, $\Gamma$.  Obviously, the plausibility will be one if $\Gamma\Rightarrow\psi$ and zero if $\Gamma\psi$ constitutes a contradiction.  The other situations are shown in the following example.

  Consider the meeting of two gamblers who have, by means unspecified, come into possession of a Stern-Gerlach magnet.  Being as they are, they decide to place wagers on measurements of a beta ($\beta^-$) decay process.  In order to decide the winner, both agree on the same method of classifying the measurement outcome.  The most apparent measurement would be whether or not the following event occurred.
\begin{quote}
A: An electron is observed in the time interval $t,t+dt$.
\end{quote}
However, they find themselves unable to assign $\Pr{A|I}$ because of a large amount of uncertainty on the physics of the experiment.  Then one of them notices that the device can tell them not only if an electron has been observed, but also if it has positive or negative spin.  This changes their prior information for the problem, since they recognize that there are now two elementary hypotheses: A, observed with spin-up, or B, observed with spin-down.  They are believed to be mutually exclusive, so that they know $(A\oplus B)$.  According to the principle of indifference,
\begin{equation}
\label{eq:indiff}
\Pr{A|(A\oplus B)I} = \frac{1}{2}
.
\end{equation}

  As their previous state of knowledge was unable to distinguish between these two events, it implicitly combined both of these two elementary hypotheses into a single event, which they held to be un-assignable.  However, it seems that the principle of indifference should have some bearing on the question of $\Pr{A|I}$, since in Aristotelian logic, $A$ must always be either true or false.  Representing this two-valued foundation of Aristotelian logic as $L$, Cox has derived the sum rule, $\Pr{A|LI}+\Pr{\bar A|LI}=1$.  In this case, assuming $L$ is equivalent to assuming $A$ and $\bar A$ are mutually exclusive events and that one must occur, a situation represented by (\ref{eq:indiff}).  In the case of Aristotelian logic, then, any reasoning on a proposition, $A$, on the left-side of $(A|J)$, must be preceded by assuming $A$ and $\bar A$ are mutually exclusive and exhaustive on the right side.  An inability to assign $\Pr{A|I}$ based only on $I$ would then amount to some system of logic that does not begin by assuming $L$.
This logical complication in part explains why the problem has not yet been directly discussed, as it requires us to reason about statements which are usually considered axiomatic using Aristotelian logic, $L$.

  Is such a system possible, and if so, does it serve any useful purpose?  Jaynes considered this relaxation to be required for reasoning about more vaguely defined propositions such as whether a defendant did or did not exercise reasonable judgement in a medical malpractice suit.
A corollary of the present question is the construction of a non-Euclidian geometry, which is indeed possible when one does away with the assumption that through one point, only one parallel may be drawn to a given straight line\cite{hpoin07}.  We argue by analogy that the product rule is more fundamental than the sum rule, and that the most important use of removing the rule ($A\oplus \bar A$) is to make explicit the assumptions on how logical propositions must inter-relate.  For example, if $A$ represents the proposition that a defendant exercised reasonable judgment, both $A$ and $\bar A$ may be held to be absolutely true, but for different choices made by the defendant.  In order to make definite conclusions, however, it will be necessary to define a set of mutually exclusive hypotheses -- for example by enumerating individual actions and measurable ethical standards.  Once a set of mutually exclusive hypotheses is defined, a problem of deciding plausibilities in the absence of this assumption may be reduced to one in Aristotelian form.

  In addition to assuming the product rule, it will be necessary to define a set of operations in a reduced Boolean algebra where the plausibility of $A\bar A$ may be nonzero.  The contradiction in this statement disappears when $\bar A$ is defined to be a new proposition, say $B$, independent of $A$ unless some prior information is present relating the two.  It would seem that by thus removing the operation of negation, a reduction to Aristotelian logic is always possible.  More precisely, the proposed system of logic, $\phi L$ contains the conjunction and disjunction in the usual sense, but not negation.  
In order to equate $\bar A$ and $B$ in the Aristotelian sense, we must then know that $A$ and $B$ are mutually exclusive and that one or the other is always true.  Because of this property, statements in $\phi L$ cannot be disproven unless some relations between them are first assumed.

  We thus add the further relations, `$\oplus$' to mean that two propositions are mutually exclusive and exhaustive, `$\Rightarrow$' to mean that the left proposition is logically equivalent to the conjunction (i.e. $A(A\Rightarrow B) \iff AB(A\Rightarrow B)$) and `$\iff$' to mean that two propositions are logically identical.  The Aristotelian expansions, $A\iff B = AB+\bar A\bar B$ and $A\oplus B = A\bar B + \bar A B$ may not hold in general, and in their place, $\Rightarrow$ and $\iff$ define the set of substitution rules which may be used.
The principle of contradiction is thus $\Pr{AB|A\oplus B}=0$.  No contradiction can be deduced without the mutual exclusivity clause, thus $\Pr{\bar A|A\iff \bar A(\phi L)}$ is undefined, whereas $\Pr{\bar A|(A\iff\bar A)(A+\bar A)}=1$.  It is also evident that $A\iff A$ is always to be assumed.  From the product rule, $\Pr{A|C}=\Pr{AA|C}=\Pr{A|AC}\Pr{A|C}$, so that the syllogism is likewise reduced to $\Pr{A|AC}=1$, irrespective of whether or not $C$ contains $\bar A$.\footnote{However, it does not make sense to admit logically contradictory prior information such as $AB(A\oplus B)$.}  At this point, all that can be said of the disjunction is that $A+A$ is equivalent to $A$, $A\Rightarrow A+B$ and that $\Pr{A+B|C} \ge \Pr{A|C}$.

  Note that $\phi L$ is consistent, since any sentence, $\psi$, in $\phi L$ can be converted to one, $\psi'$, in $L$ by symbolically re-labeling elementary propositions such that $\psi$ is true if and only if $\psi'$ is true and $\psi$ is reducible to a contradiction if and only if $\psi'$ is so.  The construction of $\psi'$ may be accomplished simply by replacing all negated elementary propositions (only individual literals may be negated in $\phi L$) with new elementary propositions.  The rules of Aristotelian logic for this sentence are in one to one correspondence with those of $\phi L$.  Note that distributing any negations for an expression in $L$ and adding to this expression an Aristotelian clause, $(A\oplus \bar A)$, for each elementary proposition that appears constitutes the reverse transformation.


  We now show that it is admissible to use the product rule to completely expand Eq.~\ref{eq:indiff} for comparison to $\Pr{A|I}$.
\begin{equation*}
\frac{1}{2} = \Pr{A|(A\oplus B)I} = \frac{\Pr{A(A\oplus B)|I}}{\Pr{A\oplus B|I}}
 = \frac{\Pr{A|I}}{\Pr{A\oplus B|I}} 
\end{equation*}
By defining a set of possible assignments, assuming $A\oplus B$ reduces statements about $A$ and/or $B$ to Aristotelian form.
From this example it is evident that the only information required to assign a probability using the principle of indifference is the number of elementary hypotheses which may be measured.  For $A(A\oplus B)$, there is only one hypothesis, and it is formally undecidable.  On the contrary, $A\oplus B$ implies that there are two possibilities, since there are two elementary hypotheses $A$ or $B$ making this expression true.  Therefore, we define the principle of indifference as one basing its determination of plausibility completely on the number of distinct truth assignments which may confirm a logical expression in $\phi L$.  In the case of only one, undecidable assignment (represented as $\varphi$), the principle of indifference gives an unknown constant.
\begin{equation}
\label{eq:const}
\Pr{A|I} \text{const.} \equiv \Pr{\varphi|I}
\end{equation}
According to the product rule, principle of indifference must therefore assign a likelihood to compound propositions as $\Pr{A\oplus B|I} = 2 \Pr{\varphi|I}$.

  In general a logical sentence, $\Gamma$, represents some assumptions on certain, contradictory, possible, and identical hypotheses.  Writing the set of literals contained by $\Gamma$ as $x_1,x_2,\ldots,x_n$, a basic set of hypotheses, $\mathbb S_\psi$, consists of the $2^n-1$ conjunctions from all possible (non-null) combinations of the $x_i$.  However, only some subset, $\Omega$, of $\mathbb S_\psi$ will be possible given $\Gamma$.  We may use the principle of indifference to assign likelihoods over this space as
\begin{equation*}
\Pr{\psi|\Gamma I} = \frac{1}{|\Omega(\Gamma)|}, \; \psi\in\Omega\subseteq\mathbb S_\psi
\end{equation*}
Next, we rigorously define $\Omega$ as the set of all conjunctions of $x_i$ (i.e. $\psi\in\mathbb S_\psi$) that raise the status of $\Gamma$ to certainty ($\psi\Rightarrow\Gamma$ so that $\Pr{\Gamma|\psi I}=1$).  
Other $\psi'$ with either be undecidable, with no relevance to $\Gamma$, or contradictory ($\Pr{\Gamma|\psi' I} = \Pr{\psi'|\Gamma I}=0$); neither will contribute to Eq.~\ref{eq:LR}.
Since each literal is represented as a unique $x_i$, we may visualize the set of conjunctions in the usual sense of a truth table, where false is taken to mean `not present.'  The product rule is now sufficient to show that
\begin{align*}
\Pr{\psi|\Gamma I} &= \frac{1}{|\Omega|}, \; \psi\in\Omega(\Gamma) \\
                   &= \frac{\Pr{\Gamma\psi|I}}{\Pr{\Gamma|I}} = \frac{\Pr{\Gamma|\psi I}\Pr{\psi|I}}{\Pr{\Gamma|I}} \\
                   &= \frac{\Pr{\varphi|I}}{\Pr{\Gamma|I}} \\
\Rightarrow & \Pr{\Gamma|I} = |\Omega| \Pr{\varphi|I}
.
\end{align*}
We note that $\Omega$ often takes the form of a product space, {\em e.g.} for \\
$\Gamma = (\oplus(A,B,C,\ldots)) (\oplus(A',B',C',\ldots)) (\cdots)$.  In this expression, we have defined $\oplus(\cdot)$ as expanding to a conjunction of exclusive-ors on all pairwise combinations in $\cdot$, so that only one expression from the argument set may be true at once.
When $\Omega$ is explicitly present as an assumption, then we may define each element of $\Omega$ as an elementary state to give the Komologrov system of probability, in which the sum rule,
\begin{equation}
\Pr{\Omega'|\Omega I} = \sum_{\psi\in\Omega'} \Pr{\psi|\Omega I} = \sum_{\psi\in\Omega'} \frac{\Pr{\psi|I}}{\Pr{\Omega|I}},\; \Omega'\subseteq \Omega
,
\end{equation}
becomes valid.
However, it should be noted that the set $\Omega$ is not itself elementary, but instead constructed from elementary hypotheses of the form `$x_1$ is true,' etc.  If the prior information, $\Gamma$, does not state that $x_1$ and $x_2$ are mutually exclusive, for example if $\Gamma=x_1+x_2=(x_1+x_2)(x_1+x_2)=x_1+x_2+x_1x_2$, then $\Omega = \{x_1, x_2, x_1x_2\}$.  In the present paper, we use $\Omega$ instead of $\Gamma$, since we have not proved that the reverse mapping $\Omega :\to \Gamma$ is unique.

\bibliographystyle{spmpsci}

\end{document}